\documentclass[12pt]{article}
\usepackage[compat=1.1.0]{tikz-feynman}
\usepackage{slashed}
\usepackage{caption}
\usepackage{amsmath}
\usepackage{amssymb}
\usepackage{braket}
\usepackage{breqn}
\usepackage[]{graphicx}
\usepackage{subfig}
\usepackage{xcolor}
\usepackage[margin=2.0cm]{geometry}

\usepackage[
    backend=biber,
    style=numeric-comp,
    sorting=none,
    sortlocale=de_DE,
    natbib=true,
    url=true, 
    doi=true,
    eprint=true
]{biblatex}

\usepackage[]{hyperref}
\hypersetup{
    colorlinks=true,
}

\usepackage{authblk}
\usepackage{slashed}

\begin{document}

\title{\textbf{Resonance refraction and neutrino oscillations}}

\author[1,3] {Alexei Y. Smirnov}
\author[2,3] {Victor B. Valera}
\affil[1]{Max-Planck-Institut f\"ur  Kernphysik,  69117  Heidelberg,  Germany}
\affil[2]{Niels Bohr International Academy, Niels Bohr Insitute, University of Copenhagen, 
DK-2100 Copenhagen, Denmark}
\affil[3]{Abdus Salam International Centre for Theoretical Physics, Strada Costiera 11, 
34014 Trieste, Italy}

\setcounter{Maxaffil}{0}
\renewcommand\Affilfont{\itshape\small}

\date{\today}
\maketitle

\begin{abstract}

The refraction index and matter potential 
depend on neutrino energy and this dependence
has a resonance character associated to the production of the mediator
in the $s-$channel. For light mediators and light particles
of medium (background) the resonance can be realized at energies accessible 
to laboratory experiments. We study properties of the energy dependence  of the potential  for
different C-asymmetries of background. Interplay
of the background potential and the vacuum term leads to (i) bump in the oscillation
probability in the resonance region, (ii) dip 
related to the MSW resonance in the background,
(iii) substantial deviation of the effective $\Delta m^2$ above the resonance 
from the low energy value, {\it etc.} 
We considered  generation of mixing in the background.
Interactions with  background  shifts the  energy of usual MSW resonance  
and produces new MSW resonances.  
Searches of the background effects allow us to put bounds on 
new interactions of neutrinos and properties of the background.
We show that explanation of the MiniBooNE excess, as the 
bump due to resonance refraction,  is excluded.

\end{abstract}
\newpage

\section{Introduction}

The Wolfenstein potential\footnote{In what follows we will consider 
potentials which are related to the refraction index
$n$ as $V = (n - 1)p$, where $p$ is the momentum of
neutrino.},  which describes the matter effect on
neutrino oscillations, do not depend on the neutrino energy
\cite{Wolfenstein-PhysRevD,Opher:1974drq,Barger:1980tf,Langacker:1982ih}.  
This is the consequence of

(i) large  mass of mediators of interactions, $M_{med}$, 
or low energies of neutrinos, so that the total energy
in the CMS: $\sqrt{s} \ll M_{med}$.
Recall that originally the potentials were derived using the
4 fermion point-like interactions.

(ii) the C- (CP-) asymmetry of background. In the C-symmetric medium
in the lowest order the potentials are zero. 

In general (independently of the C-asymmetry)   
substantial dependence of the potentials on energy
should show up at  energies $\sqrt{s} \simeq M_{med}$. Furthermore,
exchange of mediator in the $s$-channel leads to the resonance 
character of this dependence \cite{LUNARDINI2000260}. 
We will call this phenomenon the {\it resonance refraction}.

In the Standard Model the mediators of neutrino interactions are
$W$,  $Z^0$ as well as  $H^0$. $Z^0$ leads to the resonance refraction in
the $\bar{\nu} \nu - $ annihilation.
In resonance the potential is exactly zero 
and changes the sign with
energy change. Above the resonance energy the potential 
has $1/E$ dependence similar to  the usual kinetic
term related to mass squared difference \cite{LUNARDINI2000260}. 
In principle, this refraction can be realized
in scattering on the ultra high energy 
cosmic neutrinos on relic neutrino background ($E \geq 10^{21}$ eV
in the present epoch) \cite{LUNARDINI2000260}.
The $W-$boson exchange  produces the resonance refraction
in the $\bar{\nu}_e e-$ scattering, {\it i.e.},  in the Glashow resonance.
For electrons at rest this requires the neutrino energy $\sim 6.4$ PeV.
We comment on possibility of observational effects in sect. 3.8.

For light mediators and light scatterers
(their existence implies  physics beyond the SM) the resonance refraction
can be realized at low energies accessible to existing experiments.
The resonance refraction leads to increase of the oscillation phase
which can  dominate over the vacuum phase in the energy range around
the resonance. This produces an enhancement of the oscillation effect
which would be negligible without resonance refraction.
Such an enhancement was used in \cite{Asaadi_2018} to explain the low energy excess of 
the MiniBooNE events \citep{collaboration2020updated}. In this explanation the medium was 
composed of the overdense relic neutrinos.

Potentials induced by light mediator in medium with light scatterers were  computed 
recently in connection to possible existence 
of light dark sector and light dark matter  
\cite{Nieves:2018vxl,Nieves:2018ewk,Ge:2018uhz,Choi:2019zxy,Babu:2019iml,Choi:2020ydp,Ge:2019tdi}.  
Mediators  and scatterers of different nature were explored:
fermions, scalars,  gauge bosons.
Various bounds on couplings of neutrinos with new light sector
were obtained 
\cite{Berlin:2016woy,Rodejohann:2017vup,Lindner:2018kjo,Arcadi:2019uif,Lindner:2016wff,Farzan:2018gtr,
Brdar:2018qqj,Bjorken:2009mm,Batell:2009di,Essig:2010gu, Lees:2014xha,TheBABAR:2016rlg,Harnik:2012ni,
Adelberger:2006dh,Schlamminger:2007ht,Choi:2019ixb}.

In this paper we focus on phenomenon  of resonance refraction itself
presenting results in a model independent way. We study in detail dependence
of the resonant potentials on energy for different values of
the $C-$ asymmetry of background. We consider interplay
of resonance potentials with usual vacuum (kinetic) term as well with usual
matter potential. New interesting features are realized, such as shift of
the usual MSW resonances,  increase or decrease of the effective
mass squared difference with energy, {\it  etc}.
We identify signatures of the resonance refraction and outline  
possible observable effects.
As an illustration,  we apply our results to the MiniBooNE excess and show that
explanation  \cite{Asaadi_2018} is excluded.
In general,  applications can include explanations of some energy localized anomalies. 
In the absence of anomalies the bounds can be established 
on background parameters (densities, characteristics
of scatterers) and neutrino couplings.

The paper is organized as follows. In sect. 2 we introduce
interactions of neutrinos with new
light sector. 
We compute potentials due these interactions 
and study resonances in these potentials.
In sect. 3 we discuss effects of interplay of the background potential 
with vacuum (kinetic) term and usual matter potential.
We consider possible observational effects and applications of the results, in particular, 
to an explanation of the MiniBooNE excess in sect. 4. Conclusions follow in sect. 5.

 

\section{Potentials and  resonances}
\label{Sec: TheoryRR}

\subsection{Neutrino interactions with new light sector}

In this paper we focus on phenomenon of resonance refraction itself,
and present our results in general and universal form
valid for different mediators and particles of background. 
We consider the simplest (minimal) light sector composed
of new scalar $\phi$ (which can be
real or complex) with mass $m_\phi$ and fermion 
$\chi$  with mass $m_\chi$. We comment on some  extensions of this sector later. 
Interactions of the SM neutrino
mass states $\nu_{iL}$  ($i = 1, 2, 3$) with these new particles are described by 
\begin{equation}
\label{Eq:LagrangianGrl}
    \mathcal{L}^{NSI}  \supset g_i \Bar{\chi}  \nu_{iL} \phi^* + {\rm h.c.}
\end{equation}
$\phi$ may acquire VEV, thus contributing to neutrino mass.
Then for  single $\chi$  only one neutrino (combination of $\nu_i$)
will acquire mass by VEV of $\phi$. 
We assume that some other sources of $\chi$ and 
neutrino masses exist, e.g. the see-saw mechanism,
so that $\chi$ and all $\nu_i$ acquire  different masses 
and in general these masses are not related to $g_i$.


As an option several new fermions $\chi_j$ can be introduced.
Notice that $\chi_i$ themselves can be 4 component Dirac particles
which implies more degrees of freedom.
$\chi_R$ can be the left antineutrino, so that
neutrinos are Majorana particles.

The coupling can be generated via mixing
of singlet scalar field
$\phi$ with the Higgs boson doublet (Higgs portal) \cite{Khan:2017ygl}.
Alternatively,  $\phi$  can couple
with RH singlet (sterile) neutrino, which in turn, couples
(mixes) with active neutrinos (lepton and Higgs doublets) -
that is, via the RH neutrino portal.
In the Majorana case the singlet $\phi$ should mix with the neutral component
of the Higgs triplet.


The couplings (\ref{Eq:LagrangianGrl}) were considered
in various contexts before 
\cite{Rodejohann:2017vup,Lindner:2016wff,Farzan:2018gtr,Brdar:2018qqj,Batell:2009di,Essig:2010gu,Harnik:2012ni,Schlamminger:2007ht}.
For light new particles $m_\phi, m_\chi \ll 1$ GeV, a number of generic
bounds were obtained.
The bounds are based on possible transitions
$\nu \rightarrow \chi + \phi$.

Notice that refraction is induced by the elastic forward scattering
being proportional to $g^2/ M_{med}^2$. Therefore it does not disappear
in the limit $g \rightarrow 0$, provided that $M_{med}$
decreases in the same way as $g$. This allows us to avoid most
of the bound based the inelastic processes for which
$\sigma \propto (g^2/q^2)^2$, and the transfer momentum
squared $q^2$ is restricted from below by condition of
observability.

The  laboratory bounds on $g$ are rather weak:
$ g_\phi \lesssim 10^{-3}$ 
for masses $m_\phi < m_K$ ($K-$ meson mass).
They  follow, in particular,  from additional
contribution to the decay $K \rightarrow \mu \chi \phi$.
Much stronger bounds follow  from  Cosmology (BBN, CMB data, structure
formation) and astrophysics (star cooling, supernova dynamics and SN87A
neutrino observations).
They give the bound
\begin{equation}
g_\phi \lesssim 10^{- 7}.
\label{eq:gbound}
\end{equation}

Elastic forward scattering due to the interactions  (\ref{Eq:LagrangianGrl})
produces  the effective potentials $V_i$  for neutrino
mass states in medium.
There are two possibilities even for simplest
case of (\ref{Eq:LagrangianGrl})
(i) $\phi$ plays the role of mediator while $\chi$ form a background, and
{\it vice versa}: (ii) $\chi$ is
the  mediator while $\phi$ form a background.

\subsection{Potentials in the fermionic background}

We consider first the case of strong hierarchy of couplings: $g_3 \gg g_2, g_1$, 
so that $\nu_3$ couples with background, while interactions of others can be neglected.  
In this case the interactions, and consequently the potentials,  are diagonal 
in the mass basis. We will discuss couplings of  all three neutrinos later in sect. 3.7.  
Also we comment on the case of three $\chi_j$.

\begin{figure}
	\centering
	\includegraphics[width=\textwidth]{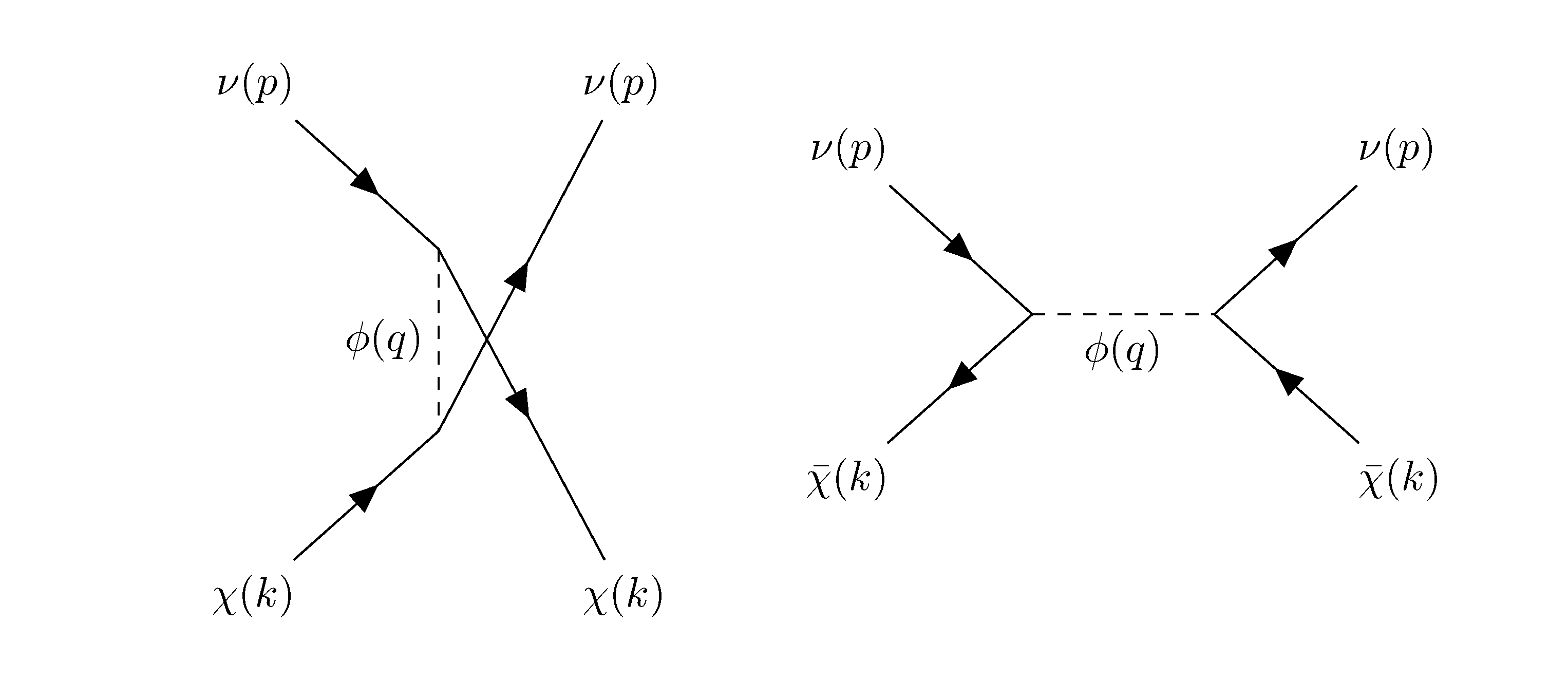}
	\caption{Feynman diagrams for scattering of neutrinos on a 
		background composed of fermions 
		$\chi$ (left) and antifermions $\bar{\chi}$ (right).}
	\label{fig: Diagram1}
\end{figure}

We consider background  composed of fermions $\chi$ 
and antifermions $\bar{\chi}$ with number  densities  
$n_{\chi}$ and  $\bar{n}_{\chi}$ correspondingly. 
The C-asymmetry of the background can be defined as  
\begin{equation}
\epsilon \equiv \frac{n_\chi - \bar{n}_\chi}{n_\chi + \bar{n}_\chi}.
\label{eq:asymm}
\end{equation}
The mediator is a scalar $\phi$ and 
the diagrams of the neutrino scattering on $\chi$ (left)  and   
$\bar{\chi}$ (right) are shown in Figure \ref{fig: Diagram1}.  
For $m_\phi > m_\nu, \, m_\chi$ the right diagram  
with  the  s-channel exchange  produces  resonance.  

To obtain the potential, we integrate the 
matrix element of the process 
over the momentum of particle $\chi$ with distribution  
function, $F_\chi(k)$. The latter is normalized as
\begin{equation}
\label{Eq: Dist_norm}
    \int d^3 k F_\chi(k) = n_\chi,
\end{equation}
and similarly for $\bar{\chi}$. 
The left ($u$-channel) and right ($s$-channel) diagrams in Fig. \ref{fig: Diagram1} give correspondingly 
the potentials 
\begin{eqnarray}
    V_{ui} & = & \int d^3 k F_\chi(k) \bra{\nu_{i, p} 
\chi_k} g^\dagger_i \bar{\nu}_i P_R \chi 
\frac{1}{q^2 - m^2_\phi} g_i \bar{\chi} P_L \nu_i \ket{\nu_{i,p} \chi_k},
\label{eq:v1i}\\
V_{si} & = & \int d^3 k F_{\bar{\chi}}(k) \bra{\nu_{i,p} \bar{\chi}_k} 
g^\dagger_i \bar{\nu}_i P_R \chi 
\frac{1}{q^2 - m^2_\phi + im_\phi \Gamma_\phi} 
g_i \bar{\chi} P_L \nu_i \ket{\nu_{i,p} \bar{\chi}_k},
\label{eq:v2i}
\end{eqnarray}
Here in the propagator we added the term with total width of $\phi$. In vacuum  
\begin{equation}
\Gamma_\phi^0 = \sum_i \frac{g_i^2}{8\pi}m_\phi \approx \frac{g^2}{8\pi}m_\phi ,  
\label{eq:width}
\end{equation} 
($\phi \rightarrow \nu \, \chi$), 
where $g = g_3$ and we take $g_1 = g_2 = 0$. 
In medium the term with $\Gamma$ is modified (see below). 

We assume first that the background particles 
are at rest which is valid  for cold gases like dark matter (DM) 
or relic neutrinos from the cosmological neutrino background (C$\nu$B). 
Then  $F_\chi (k) = n \delta(\vec{k})$ 
and the integrals in (\ref{eq:v1i}) and (\ref{eq:v2i}) give
the total potential 
\begin{equation}
\label{Eq:PotentialScalarMed}
V^B \equiv  V_{u } + V_{s} 
= \frac{|g|^2}{2} \left[\frac{n_{\chi}}{(2E_\nu m_\chi + m^2_\phi)} + 
\frac{\bar{n}_{\chi} 
(2E_\nu m_\chi - m^2_\phi) }{(2E_\nu m_\chi - m^2_\phi)^2 
+ (m_\phi \Gamma_\phi)^2} \right]. 
\end{equation}
This expression differs from expression for potential in \cite{Asaadi_2018},  
but coincides with that in \cite{Choi:2019zxy}.

We obtain similar result for moving  
$\chi$ with the only substitution $m_\chi \rightarrow E_\chi$, 
if  the  angular distribution is isotropic.  This is important for the degenerate gas 
with large overdensity when  the Fermi momentum $p_f \gg m_\chi$.

The second term in (\ref{Eq:PotentialScalarMed}) has a  resonance  dependence on  energy 
(pole of propagator) with the resonance energy  
\begin{equation}
\label{Eq:resener}
E_R \equiv \frac{m^2_\phi  - m_\chi^2 - m_\nu^2}{2 m_\chi} \approx 
\frac{m^2_\phi}{2 m_\chi}. 
\end{equation}
At $E_R$  the contribution $V_s$ is exactly zero and 
it changes the sign with energy change. 
The amplitude of scattering becomes purely imaginary,  
which corresponds to production of the on shell $\phi$. 
In terms of the resonance energy the potential (\ref{Eq:PotentialScalarMed}) 
can be rewritten as  
\begin{equation}
\label{Eq: PotentialScalarParam}
V^B =  \frac{|g|^2 (n_\chi + \bar{n}_\chi)}{8 m_\chi} 
\left[ \frac{(E - E_R) (1 - \epsilon)}{(E - E_R)^2 + (\xi E_R)^2} 
+ \frac{1 + \epsilon}{E + E_R } \right],
\end{equation}
where 
\begin{equation}
\label{eq:xidef}
\xi \equiv \frac{\Gamma_\phi}{m_\phi},
\end{equation}
and in vacuum 
\begin{equation}
\label{eq:xidef0}
\xi^0 \equiv \frac{g^2}{8\pi}.   
\end{equation}

Let us introduce  a 
dimensionless parameter
\begin{equation}
y \equiv \frac{E}{E_R}. 
\end{equation}
In terms of $y$ the expression for the potential (\ref{Eq: PotentialScalarParam}) 
becomes 
\begin{equation}
\label{Eq:poten-dim}
V^B = \frac{1}{2} V_0^B \left[ \frac{(1 - \epsilon)(y - 1)}{(y - 1)^2 + \xi^2}
+ \frac{1 + \epsilon}{y + 1} \right], 
\end{equation}
where 
\begin{equation}
\label{eq:vzero}
V_0 \equiv \frac{g^2}{2m_\phi^2} (n_\chi + \bar{n}_\chi).  
\end{equation} 
In this way we can disentangle dependencies of the potential on 
relevant physical quantities: $V_0$ depends on parameters of mediator,  
$g$ and $m_\phi$,  and on total density of scatterers in a background. 
It has a form of the standard matter potential 
at low energies with $G_\phi = g^2/2 m_\phi^2$.  
The parameter $\xi$ is proportional to the coupling  constant squared,  
while the mass of $\chi$ enters via $E_R$.  
$V_0$ is introduced in such a way that for $y \rightarrow 0$ 
we have $V^B \rightarrow \epsilon V_0$,  and consequently,  for  $\epsilon =  \pm 1$:  
$V^B = \pm V_0$, thus reproducing
the standard Wolfenstein potential.

\subsection{Potentials in the bosonic background}

For the scalar particle  background and fermionic mediator 
the potential  is similar to the one computed before. 
In the lowest order in $g^2$, up to factor of 2  the potential has 
the same expression as in (\ref{Eq:PotentialScalarMed}) with the following
substitutions
$$
m_\phi \leftrightarrow m_\chi, \, \, \,
n_\chi \rightarrow n_\phi,  \, \, \,
\Gamma_\phi \rightarrow \Gamma_\chi. 
$$
Thus,
\begin{equation}
V^\phi \approx 2 V^\chi
(m_\phi \rightarrow m_\chi, \,
m_\chi \rightarrow m_\phi,
n_\chi \rightarrow n_\phi, \,
\Gamma_\phi \rightarrow \Gamma_\chi).
\label{potphi}
\end{equation}

The resonance is realized if $m_\chi > m_\nu + m_\phi$,
and the resonance energy equals
\begin{equation}
E_R \simeq  \frac{m_\chi^2}{2 m_\phi}.
\label{ereschi}
\end{equation}
In terms of resonance energy the potential can be written
in  exactly the same form as in  (\ref{Eq:poten-dim}) with
\begin{equation}
V_0^\phi = \frac{g^2}{2 m_\chi^2} (n_\phi + \bar{n}_\phi), 
\label{v0phi}
\end{equation}
and
\begin{equation}
\epsilon^\phi  \equiv
\frac{n_\phi - \bar{n}_\phi}{n_\phi + \bar{n}_\phi}.
\label{epsilonphi}
\end{equation}

The difference from the fermionic background case may appear in higher orders in $g^2$ 
due to fermionic nature of mediator $\chi$. Now the amplitude of scattering is proportional to  
$\slashed{q} = \slashed{p} + \slashed{k}$: 
$A =  \slashed{p} \Sigma_\nu + \slashed{k}\Sigma_\chi$. The first term gives contribution 
to renormalization 
of the wave function of neutrino:   
$\nu  = (1 + \Sigma_\nu/2) \nu'_L$, while the second one  generates the potential: 
for the background at rest $\gamma^0 m_\phi \Sigma_\chi = \gamma^0 V $.  
Renormalization leads to change of the potential: 
$V' =  (1 + \Sigma_\nu^*/2) V (1 + \Sigma_\nu/2) = V(1 + \Sigma_\nu)$ 
(as well as usual kinetic term) \cite{Choi:2019zxy}.    
The correction is of the order $g^2$. 
In this order one should take into account also loop corrections to external neutrino lines 
All these corrections have the same nature 
and can be described  by tree level diagrams with multiple scattering on a background: 
$\nu + \phi^*   \rightarrow \chi  \rightarrow  \nu + \phi^*,  \nu + \phi^*   \rightarrow \chi ... $. 
Alternatively it can be treated as resummation of self-energy loop diagrams. 
The high order corrections will not change general properties (energy dependence) of 
potentials.  In the lowest order properties of the resonances in the scalar
and fermion backgrounds are the same. The difference 
appears in applications and implications for theory.

\subsection{Resonance, energy smearing, coherence}

In resonance, $y = 1$, the $s-$component of the potential (\ref{Eq:poten-dim})
is zero for any asymmetry, $V_s = 0$,  and only non-resonance component contributes.
The potential has extrema at $y = 1 \pm \xi$: 
\begin{equation}
|V^{\rm max}| = \frac{V_0}{4} \left(\frac{1 - \epsilon}{\xi} + 1 + \epsilon \right) 
\approx \frac{V_0}{4} \frac{1 - \epsilon}{\xi} . 
\label{eq:resvmax}
\end{equation}
So, in resonance the  enhancement is given by inverse coupling constant squared. 
The energy interval between two extrema equals $2\xi E_R$. In these points 
the ratio of the resonant to non-resonant part equals 
\begin{equation}
\label{eq:resnonres}
\frac{V_s}{V_u} =  \frac{1}{\xi}~\frac{1 - \epsilon}{1 + \epsilon}. 
\end{equation}
Zero of the total potential is shifted with respect to $y = 1$  due to the non-resonant 
contribution as  
$$
y_0 = 1 - \frac{1}{2} \xi^2~ \frac{1 + \epsilon}{1 - \epsilon}.  
$$
The width of the peak at the half of height, 
$V_s (y_{1/2}) = 0.5 V_s^{\rm max}$, equals 
\begin{equation}
\label{eq:resvw}
|y_{1/2} -  1| = (2 + \sqrt{3})\xi \approx  3.73 \xi .
\end{equation}

For values of couplings (\ref{eq:gbound}),
$\xi  \sim \xi^0 < 10^{-15}$,  the characteristics of resonance 
in (\ref{eq:resvmax})  - (\ref{eq:resvw})
(width and enhancement in the peak) have no physical sense.
One should take into account (i) smearing of the peaks due to integration 
with  distribution of the background $\chi$ over momenta, 
which differ from $\delta$ function, (ii) averaging over uncertainty in neutrino energy,   
(iii) effect of density correction to the width of $\phi$, (iv) dumping due to resonance absorption.

Let $\sigma_y$ be the scale of smearing in variable $y$. The smearing leads to 
decrease of heights of the peaks and their widening. If $\sigma_y \gg \xi$, we can neglect $\xi^2$ in 
(\ref{Eq:poten-dim}). Then the height of the peak after averaging can be estimated as  
\begin{equation}
|V^{\rm max}| =   V^B(1 + \sigma_y) =    V_0 (1 - \epsilon) \frac{1}{2 \sigma_y}.  
\label{eq:vmaxsm}
\end{equation}
So that the enhancement factor is given by $1/\sigma_y$. 
The maxima shift to $y \approx \sigma_y/2$. 
Let us consider possible origings of $\sigma_y$.

The quantity $\sigma_y$ can be the  width  of $F_\chi(k)$ distribution.
Recall that deriving the potential (\ref{Eq: PotentialScalarParam}) we assumed that
the background particles are at rest,  $k_\chi = 0$.
(This can be still a possibility for condensate of scalar DM). 
For fermions $F(k)$ is not the $\delta-$function, 
but distribution with finite width. 
In Eq. (\ref{Eq:PotentialScalarMed}) one should use
(even for isotropic background) 
$E_\chi = \sqrt{m_\chi^2 + k_\chi^2}$.  
Near the resonance 
$$
\sigma_y \approx \frac{\Delta E_\chi}{ E_\chi}, 
$$
and for non-relativistic background
$E_\chi \sim m_\chi + k^2/2m_\chi$, so that 
$\Delta E_\chi \approx \Delta (k^2)/ 2 m_\chi$. 
For thermal background with temperature $T$  we can take 
$\Delta k^2 = (3T)^2$, and therefore
\begin{equation}
\sigma_y = \frac{\Delta (k^2)}{2 m_\chi^2} \approx
\frac{9 T^2}{2m_\chi^2}. 
\end{equation}
If $T = 1.945$ K and $m_\chi = 0.05$ eV,  
we obtain  the  value of enhancement $1/\sigma_y \sim 10^{4}$. 

Further smearing of the dependence of potential on energy is due to
neutrino energy uncertainty $\sigma_E$ in the oscillation
setup. In this case  $\sigma_y = \sigma_E/E_R$.

For very narrow resonance one needs to take into account the medium corrections to the 
$\phi-$propagator. The main correction is given by  the loop diagram 
$\phi \rightarrow \nu +  \chi^*  \rightarrow \phi$
with the $\chi$ propagator in a finite density medium. This medium correction corresponds to scattering 
of $\phi$ on particles of medium via neutrino as mediator: 
$\phi + \chi  \rightarrow  \nu \rightarrow   \phi + \chi$. So, whole the process consists 
of the 
transitions: $\nu +  \chi^*  \rightarrow \phi$, $\phi + \chi  \rightarrow  \nu$, 
$\nu +  \chi^*  \rightarrow \phi$, 
$\phi \rightarrow  \chi +   \nu$. 
These transitions can be treated as the induced decay of $\phi$ in medium. 
The polarization operator equals 
$$
\Pi = g^2 \frac{n_\chi}{4 m_\chi}, 
$$
which should be compared with $m_\phi \Gamma_\phi^0 = g^2 m_\phi^2/8\pi$. 
Therefore the width can be written as 
\begin{equation}
\Gamma_\phi =  \Gamma_\phi^0 \left(1 +  \frac{2\pi n_\chi}{m_\phi^2 m_\chi} \right).  
\end{equation}

Ratio of the polarization operator and   $m_\phi^2$ (the denominator outside the resonance): 
\begin{equation}
 \beta \equiv \frac{\Pi}{m_\phi^2}  =  
\frac{g^2 n_\chi}{4 m_\phi^2 m_\chi} . 
\label{eq:alpha}
\end{equation}
can be considered as the expansion parameter of the perturbation theory.

Refraction implies coherence: zero transfer momentum by neutrinos,  
and consequently, the unchanged state  of medium $|M \rangle$: $\langle M' | M \rangle \approx 1$. 
In the resonance region  
(in the s-channel) $\nu$ interacting with $\chi$ in some point $x$ 
produces nearly on-shell $\phi$
which propagates for some distance and then decays back into $\nu$ and $\chi$. So, 
the particle of medium reappears in different space-time point $x'$. 
Then the  coherence condition requires  
$\langle \chi(x') | \chi(x) \rangle \approx 1$. 
That is,  the corresponding wave functions of $\chi$ before and after 
scattering should nearly coincide. 

The time of propagation of $\phi$ between the production and annihilation 
is determined  by the  decay rate $\tau_\phi = 1/\Gamma$. 
Taking into account the Lorentz factor $\gamma = E_\phi/m_\phi$ we find 
the distance of propagation of $\phi$ in the rest frame of background:
\begin{equation}
d \approx c t_\phi = \tau_\phi  \frac{E_\phi}{m_\phi} = 
\frac{2 \pi E_\phi }{|g|^2m^2_\phi}. 
\label{eq:tr-dist}
\end{equation}
For light background particles  
the total energy of  mediator is 
$E_\phi = E + m_\chi \approx E_\nu$, 
Using the resonance condition,  
$m_\phi^2 = 2 E m_\chi$, we can rewrite 
(\ref{eq:tr-dist}) as 
\begin{equation}
d  = \frac{\pi}{|g|^2 m_\chi} \approx 6.2 \cdot 10^{9}\text{cm}
\left(\frac{|g|}{10^{-7}}\right)^{-2} 
\left(\frac{m_\chi}{1 \text{ eV}}\right)^{-1}.
\end{equation} 
$d$ should be smaller than the uncertainty 
in the position (localization) of the background particle $\Delta x \approx 1/\Delta  p_\chi$.  
This gives  the coherence condition
\begin{equation}
	d < \frac{1}{\Delta p_\chi},
\end{equation}
which imposes the upper bound on the uncertainty
\begin{equation}
\Delta p_\chi \lesssim 3 
\left(\frac{|g|}{10^{-7}}\right)^{2} 
\left(\frac{m_\chi}{1 \text{ eV}}\right) 10^{-15} \, \text{eV.}
\end{equation}

However, for a given neutrino energy most of the particles of a background 
are not  in resonance exactly and  produced $\phi$
will be out of mass shell. The virtuality can be estimated as 
$$
\Delta q \sim   \sqrt{s - m_\phi^2}  = 
\sqrt{E_R \frac{k^2}{m_\chi}}  =  \frac{m_\phi k}{m_\chi}.  
$$
Consequently, typical distance of travel is 
$d \sim m_\chi/m_\phi k $.
The  scale of localization is about $1 /n_\chi^{1/3}$. 
So,  the condition for coherence can be written as 
\begin{equation}
\frac{m_\chi}{m_\phi \langle k \rangle} \ll \frac{1}{n_\chi^{1/3}}. 
\label{eq:cohcoh}
\end{equation}

\subsection{Properties of resonance and total potential}

\begin{figure}
        \centering
        \includegraphics[width=\textwidth]{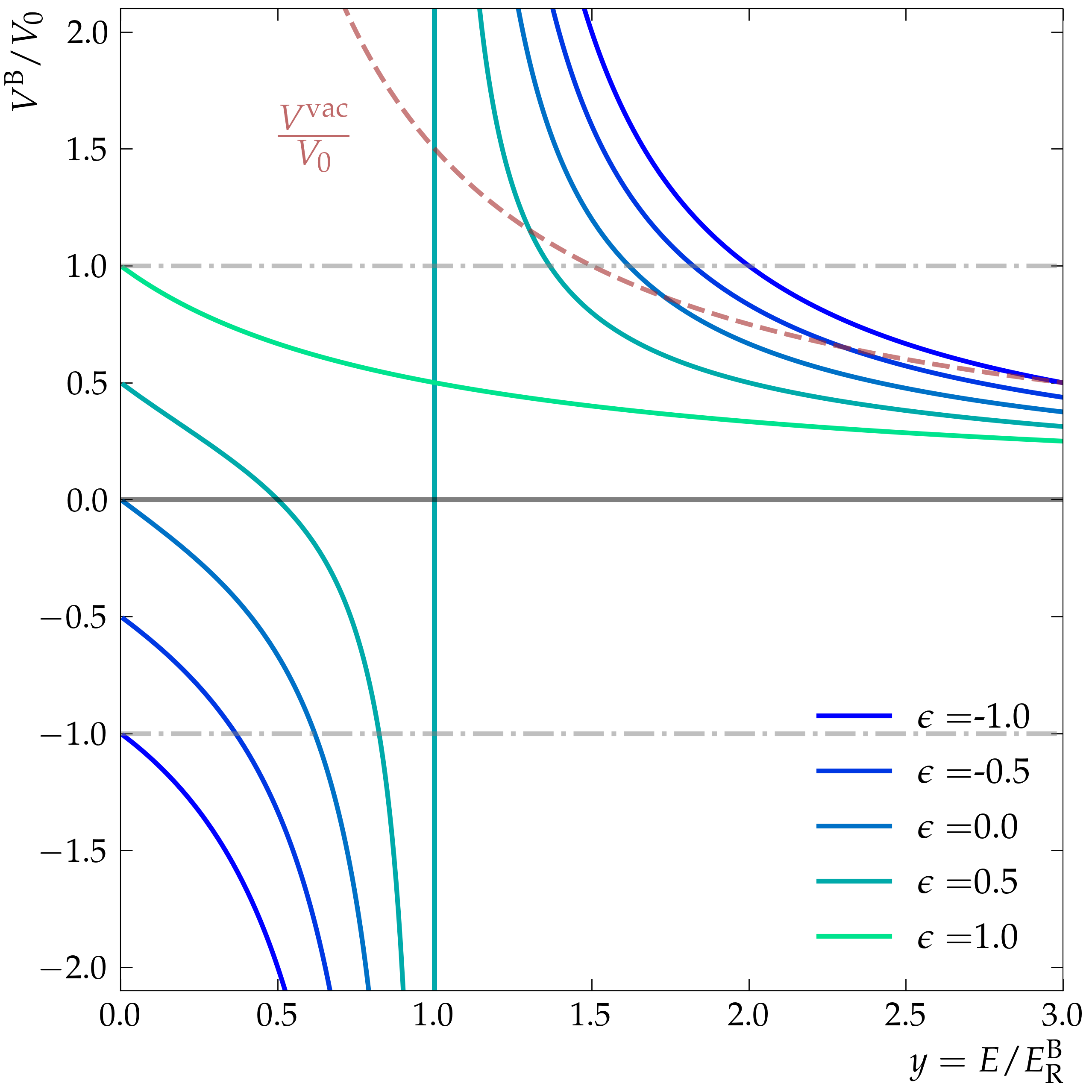}
        \caption{The dependence of the potential $V^B/V^0$ on energy, $y = E/E_R$,   
for different values of $\epsilon$. Shown also the vacuum kinetic energy 
$V^{\rm vac}/V_0$ as function of $y$. }
        \label{fig:vb-y}
\end{figure}

Outside the resonance, $|y - 1| \gg \xi$, neglecting $\xi$ 
we obtain from (\ref{Eq:poten-dim}) 
\begin{equation}
\label{eq:farfr}
V^B(y, \epsilon) = V_0~ \frac{y - \epsilon}{y^2 - 1}.
\end{equation}
In Fig. \ref{fig:vb-y} we show dependence 
of $V^B/V_0$ on $y$ for different values of asymmetry $\epsilon$ ($\epsilon = - 1 \div 1$). 
For $y = 0$: 
$$
V^B  = V_0~\epsilon, 
$$
so that for symmetric background $V^B = 0$. 
Above the resonance, $y \gg 1$,  independently of $\epsilon$ 
$$
V \approx V_0 ~\frac{1}{y}. 
$$
Thus, at $E \gg E_R$  
the potential takes the form of the standard vacuum contribution 
with $1/E$ dependence. Therefore, in principle,   
the  standard neutrino oscillations can be reproduced 
(even for massless neutrinos) 
provided that 
\begin{equation}
\frac{n_\chi}{4m_\chi} \Delta |g|^2 \simeq  \Delta m^2. 
\end{equation}
(See recent discussion in  \cite{Smirnov2020},  \cite{Choi:2020ydp}).

For particular values of $\epsilon$ we have the following dependence 
on $y$ (see Fig. \ref{fig:vb-y}). 

\begin{itemize}

\item

$\epsilon = - 1$ corresponds to pure $\bar{\chi}$  background, 
and consequently,  only the resonance contribution exists:  
\begin{equation}
\label{eq:farmone}
V^B(y, -1) = V_0 \frac{1}{y - 1}.
\end{equation}
At $y = 0$: $V(0, -1) = - V_0$,  then it decreases with increase of $y$. 

With increase of $\epsilon$ the low energy part of the potential ($y < 1$) shifts up.

\item 

$\epsilon = 0$ corresponds to  symmetric background. The potential equals 
\begin{equation}
\label{eq:farzero}
V^B(y, 0) = V_0 \frac{y}{y^2 - 1}. 
\end{equation}
$V = 0$ at $y = 0$, and then  $V^B(y, 0)$ decreases linearly below the resonance: 
$$
V(y, 0) = - V_0 y. 
$$

\item 

$\epsilon > 0$: according to (\ref{eq:farfr}) 
for  $y >  \epsilon$,  $V^B$  has positive values, 
it vanishes at  $y = \epsilon$ and then becomes negative.

\item 

$\epsilon = 1$ corresponds to pure ${\chi}$  
background and resonance is absent: 
\begin{equation}
\label{eq:farone}
V^B(y, 1) = V_0 ~\frac{1}{y + 1} 
\end{equation}
describes the 
asymptotic curve 
with $V^B/V_0 = 1$ at $y = 0$. $V^B(y)/V_0$ decreases monotonously 
from 1 to $0$ at $y \rightarrow \infty$ and at $y = 1$ 
the ratio equals $0.5$.

\end{itemize}

For $\epsilon < 1$ the dependence of potential on $y$ 
has two branches.  
In the low energy branch, $y < 1$, the ratio  $V^B/V_0$ decreases from 
$\epsilon$ at $y = 0$ down to $- (1 - \epsilon)/4\xi$ 
at $y \approx 1 - \xi$, if there is no smearing,  see eq. (\ref{eq:resvmax}). 
In the high energy branch,  $y > 1$, 
we have $V^B/V_0 > 0$,  and it decreases from 
$V^B/V_0 \sim (1 - \epsilon)/4\xi$ 
at $y = 1 + \xi$ down to zero at $y \rightarrow \infty$ (without smearing). 
The two branches are connected in the range $y = 1 \pm \xi$.
The largest effect of a background is for $\epsilon = -1$. 
With increase of $\epsilon$ both branches approach 
the non-resonance curve (\ref{eq:farone}) everywhere apart from the 
region around 1:  
$$
y  = \epsilon \div \left[\frac{1}{2} + \frac{1}{2}\sqrt{1 + 4(1 - \epsilon)} \right] 
\approx \epsilon \div  (2 - \epsilon) . 
$$

\section{Resonance refraction and oscillations}

\subsection{Background versus vacuum contributions}


Let us consider an interplay of  the background $V^B$ with  
kinetic term (``vacuum potential"): 
\begin{equation}
V^{\rm vac} (E) \equiv \frac{\Delta m^2}{2E} = \frac{V^{\rm vac}_R}{y}, ~~~~
V^{\rm vac}_R \equiv \frac{\Delta m^2}{2E_R}.
\label{vvac}
\end{equation}
We can neglect the usual matter effect  if the 
refraction resonance energy is much smaller than the MSW resonance 
energy: $E_R^B \ll E_R^{\rm MSW}$. For the Earth based experiments this 
means $E_R^{B} \ll 6$ GeV, which is realized for short baseline experiments,  
such as reactor reactor neutrino experiments, 
LSND and  MiniBooNE and low energy LBL experiments, 
e.g.,  T2K.

In general, $V^{\rm vac}$ 
can be positive or negative depending on the 
mass ordering (sign of  $\Delta m^2$). The sign is relevant since 
now we have two  contribution to the phase. In the model where 
$y_i$ correlate with masses, the potentials  $V^{\rm vac}$ and $V_0$ correlate  
too,   having the same sign. 
If both $V_0$ and  $V^{\rm vac}$  are positive, the potential $V^{\rm vac}(y)$ crosses 
$V^B(y)$ at  $y > 1$ provided that $V_R^{\rm vac}/V_0 > 1/2$.

To compare the two contributions we consider the ratio 
\begin{equation}
\kappa(y)  \equiv \frac{V^B}{V^{\rm vac}} = 
r \frac{y(y - \epsilon)}{y^2 - 1}, 
\label{ratio-v}
\end{equation}
where according to (\ref{Eq:resener}), (\ref{eq:vzero}) , 
\begin{equation}
r \equiv \frac{V_0}{V^{\rm vac}_R}
= \frac{g^2(n_\chi + \bar{n}_\chi)}{2 m_\chi \Delta m^2}. 
\label{ratio-vres}
\end{equation}
The parameter $r$ determines 
the relative strength of the background effect. 
Notice that $r$  depends on the mass of particles of the background, 
but does not depend on the mass of mediator.  
More importantly,  $r$ determines the ratio of potentials for $y \rightarrow \infty$. 

Two contributions to the phase are equal  (for $r \neq 1$) at  
\begin{equation}
y_{\rm eq} = \frac{1}{2(1 - r)}\left[
-\epsilon r + \sqrt{\epsilon^2 r^2 + 4(1 - r)}\right].  
\label{y-equality}
\end{equation}
This equation  gives $y_{\rm eq} = 1/(1 - r)$ for $\epsilon = -1$, 
and $y_{\rm eq} = \sqrt{1/(1 - r)}$ for $\epsilon = 0$. With 
decrease of $r$,  as well as increase  
of $\epsilon$ the value of  $y_{\rm eq}$ approaches 1. 
For the non-resonance case ($\epsilon = 1$)  $y_{\rm eq} = 1/(r - 1)$ and the 
equality is realized when $r > 2$. 

For the low energy  branch,  $y < 1$,  an interesting feature is 
cancellation of two contributions when    
$$
V^B = - V^{\rm vac},   
$$
which corresponds to the MSW resonance on the background.  
If $r \neq - 1$ this happens at  
\begin{equation}
y_c = \frac{1}{2(1 + r)}\left[\epsilon r + 
\sqrt{\epsilon^2 r^2 + 4(1 + r)}\right],   
\label{y-cancel}
\end{equation}
so that  $y_c = 1/(1 + r)$ for $\epsilon = -1$, 
and $y_c = \sqrt{1/(1 + r)}$ for $\epsilon = 0$. 
With decrease of $r$ and $\epsilon \rightarrow 1$ 
the cancellation point approaches 1. 
Also for $\epsilon \rightarrow 1$ we find that $y_c \rightarrow 1$. 


\begin{figure}
        \centering
        \includegraphics[width=\textwidth]{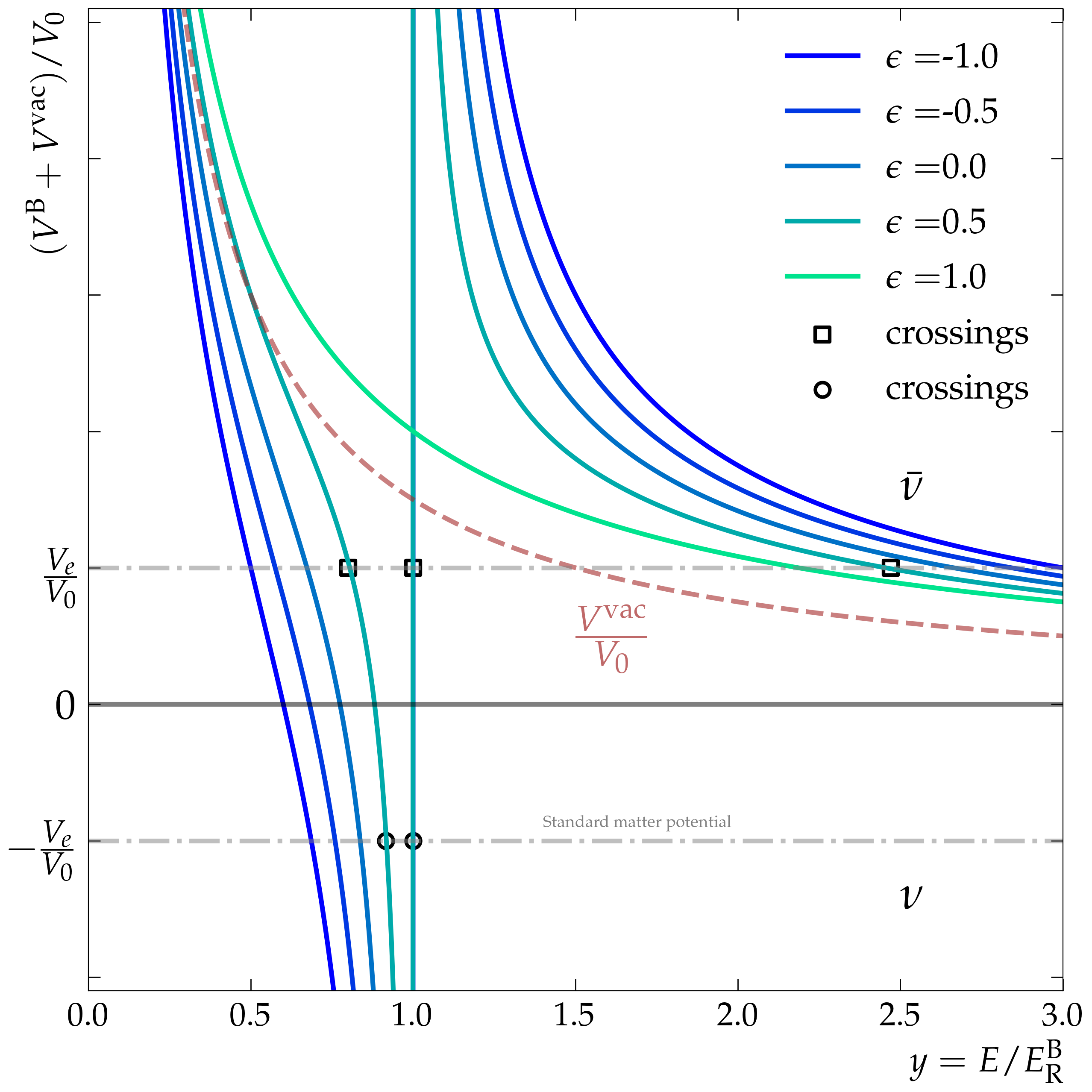}
        \caption{The dependence of the total potential,   
$(V^B + V^{\rm vac})/V_0$,  on energy $y$ for different values of $\epsilon$. 
The horizontal lines correspond to the usual matter potential $V_e/V_0$ 
for neutrinos and antineutrinos. Crossings of these lines with $(V^B + V^{\rm vac})/V_0$ 
show the points of  the MSW resonances in the neutrino (empty boxes) 
and antineutrino (empty circles) channels.}
\label{fig:vsum-y}
\end{figure}


The sum of two contributions 
\begin{equation}
V^{\rm sum} \equiv V^{\rm vac} + V^B =  
V^{\rm vac} [1 + \kappa (y)], 
\label{v-sum}
\end{equation}
in the units of $V_0$ as function of $y$, is shown in Fig. \ref{fig:vsum-y}. 
It has the following features. 
In the high energy branch $V^{\rm sum}$  increases 
from [$V^{\rm vac} (1 + r)$] at $y \rightarrow \infty$
to $V_0(1 - \epsilon)/4\xi$ at $y = 1 + \xi$ (in absence of smearing). The two contributions 
become equal at $y_{\rm eq}$ (\ref{y-equality}). 
In the low energy branch $V^{\rm sum}/V_0$ decreases from 
$V^{\rm vac}/V_0 (1 + \epsilon)$ 
at $y \rightarrow 0$, 
down to $- (1 - \epsilon)/4\xi$ at $y = 1 -\epsilon$. 
It crosses zero at $y = y_c$.  
Correspondingly, the modulus $|V^{\rm sum}|$ increases with $y$ 
at $y > y_c$ up to $V_0/\xi$. 

Thus, the background contribution distorts 
substantially 
the potential (and consequently, the vacuum phase)  dependence on $y$ in the resonance region  
$y \sim 1$: $y_c \div y_{\rm eq}$. This region shrinks with  increase of $r$ 
and $\epsilon$. Maximal distortion effect is at $\epsilon = -1$. 

\subsection{Effective mass splitting}

Effect of the  background can be treated as modification of the 
mass squared difference which depends on neutrino energy: 
\begin{equation}
\Delta m^2_{\rm eff} (y) = \Delta m^2 [1 + \kappa(y)],        
\label{dmeff1}
\end{equation}
so,  that $V^{\rm sum} = \Delta m^2_{\rm eff}(y)/2E$.  
The ratio of the effective splitting in a 
background, $\Delta m^2_{\rm eff}$, and in vacuum, $\Delta m^2$ equals
\begin{equation}
R_\Delta  \equiv \frac{\Delta m^2_{\rm eff}}{\Delta m^2} =  
\frac{V^{\rm sum}}{V^{\rm vac}} = \frac{\Phi^{\rm tot}}{\Phi^{\rm vac}}. 
\label{dmeff-ph}
\end{equation}
According to (\ref{dmeff1}) the ratio can be written as    
\begin{equation}
R_\Delta(y)  =  
1 +  r~ \frac{y(y - \epsilon)}{y^2 -1}.
\label{dmeff}
\end{equation}
For $y \rightarrow 0$ the correction disappears  
\begin{equation}
R_\Delta(y) =  1 + \epsilon r y .
\label{dmeff0}
\end{equation}
For high energies with increase of $y$ 
the ratio converges to constant value  
\begin{equation}
R_\Delta(y) =  1 + r
\label{dmeffinf}
\end{equation}
independently of $\epsilon$. 
Thus, the key consequence of interaction with background is that 
$\Delta m^2$ extracted from  data above the 
refraction resonance differs from  $\Delta m^2$  
extracted from low energy data.

\begin{figure}
        \centering
        \includegraphics[width=\textwidth]{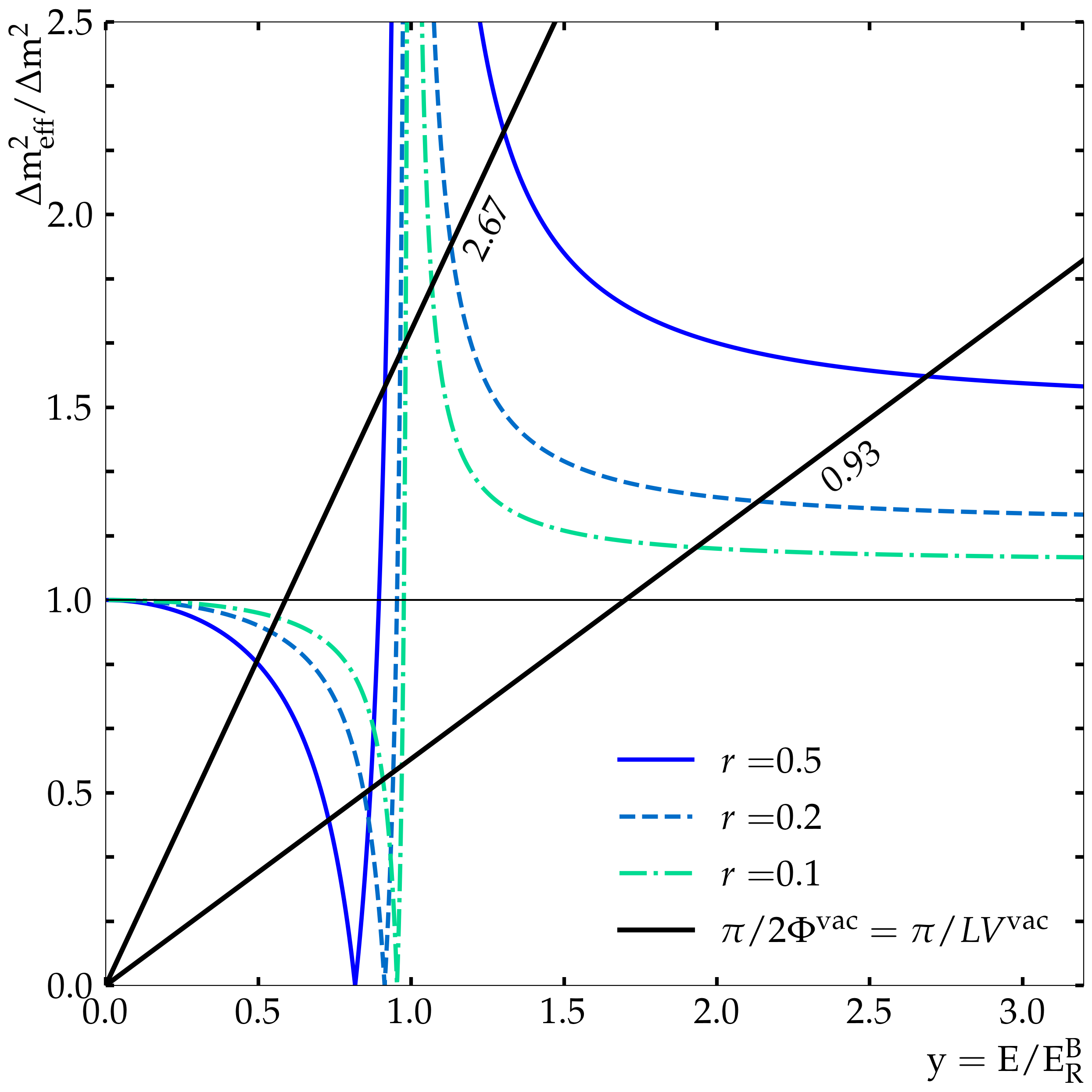}
        \caption{The effective mass splitting $\Delta m^2_{\rm eff}/\Delta m^2$ as function of $y$ 
for different values of $r$. We take $\epsilon = 0$. Shown are also lines   
$\pi/ 2L V^{\rm vac}(y)$ which correspond to two different values of baseline  
$\pi/ 2L V_R^{\rm vac}$ (numbers at the lines). Crossings of  these lines with $\Delta m^2_{\rm eff}/\Delta m^2$ 
give the points where the total phase $\Phi = \pi/2$ (see text for explanations). } 
\label{fig:dms-eff}
\end{figure}

In Fig. \ref{fig:dms-eff} we show dependence of the ratio 
(\ref{dmeff}) on $y$ for different values of $r$. 
Here the important point is $y_s$ in which  
corrections to the modulus of effective mass squared  difference  
changes the sign. It is determined by 
\begin{equation}
\left|R_\Delta \right|  = 1, 
\label{ys}
\end{equation}
or according to (\ref{dmeff1}) by $V^B(y)/V^{\rm vac}(y) = -2$. 
Solution of the corresponding equation gives
\begin{equation}
y_s = \frac{1}{2(2 + r)}\left[\epsilon r + 
\sqrt{\epsilon^2 r^2 + 8r(2 + r)}\right].   
\label{y-sign}
\end{equation}
For $\epsilon = 0$, we find $y_s = \sqrt{2/(2 + r)}$.  
Consequently,   for $r = 1.5$ it equals  $y_s = 0.87$, and for $\epsilon = -1$: 
$y_s = 0.75$. In the interval 
$y = 0 \div y_s$ the background diminishes splitting:   
$\Delta m^2_{\rm eff} < \Delta m^2$,  
and consequently, the oscillation phase. For $y > y_s$:  
$\Delta m^2_{\rm eff} > \Delta m^2$ and the phase increases. 
With decrease  of $r$ the correction decreases and 
the benchmark energies $y_c$, $y_s$ and $y_{\rm eq}$ approach 1. 
 

\subsection{Negative $\kappa$}

In the previous consideration we assumed that $\Delta m^2$ is positive, or 
more precisely, $\Delta m^2$ and  $V^B  = V^B_2 - V^B_1$ are positive simultaneously. 
That is, the potentials follow the hierarchy of masses, which is automatically 
satisfied if both differences are given by $g_2^2 - g_1^2$. 
As a consequence, $\kappa \geq 0$ and $r \geq 0$.

If, however, neutrinos have other sources of masses 
apart from VEV of $\phi$, the signs and values of $\Delta m^2$ and $V^B$ 
are independent.  
In this connection let us consider the case of negative $\kappa$ and $r$. 
Above the resonance the quantities  $V^B$ and $V^{\rm vac}$ have opposite signs. Therefore

1. The cancellation point $y_c$ (the MSW resonance on background) is 
above the refraction resonance: $y_c > 1$.

2. The ratio $R_\Delta$ increases from 1 at $y \rightarrow 0$ 
to maximum  at $y \approx 1 - \xi$ (no smearing). 

3. The dip is above the resonance peak.

4. In asymptotics, $y \rightarrow \infty$,  we have  $R_\Delta \rightarrow 1 - |r| < 1$. 
So, one expects smaller value   
of $\Delta m^2_{\rm eff}$ in comparison to the vacuum value:  $ \Delta m^2_{\rm eff} = (1 - |r|)\Delta m^2$.

\subsection{Phases and probabilities}


In the case of diagonal matrix of potentials 
in the neutrino mass basis  the background potential modifies neutrino oscillations 
via an extra contributions to the oscillation phase: 
\begin{equation}
\label{eq:phase}
\Phi  = \Phi^{\rm vac}  +  \Phi^B = 
(V^{\rm vac}  +  V^B)L, 
\end{equation}
while the mixing is unchanged. Thus,  for 
two neutrino mixing  
the  $\nu_\alpha - \nu_\beta$ transition probability equals
\begin{equation}\label{Eq: Prob}
P_{\nu_\alpha \xrightarrow{} \nu_\beta}(L,E) = \sin^2{2\theta} 
\sin^2 0.5 \Phi .  
\end{equation}
We assume here constant density of background particles. 

Since the phase $\Phi$  enters  in the observables (probability) 
as $\cos \Phi$  or $\sin^2 \Phi /2$, the change of sign of $V$ 
in the resonance 
does not lead to suppression due to integration over energy. 
(Notice that this is valid for $2\nu$ case and without matter effect. 
In the $3\nu-$ case we have interference of 
different channels with different frequencies and those terms are 
not even with respect to $V$.)

Observational effects of the background depend on the baseline 
of experiment. In Fig.  \ref{fig:dms-eff} we show the lines 
\begin{equation}
\frac{\pi}{2V^{\rm vac}L} 
= \frac{\pi}{2\Phi^{\rm vac}} = \frac{\pi y}{2V^{\rm vac}_R L}. 
\label{dmeffinf}
\end{equation}
For fixed $L$ the lines  correspond to the inverse 
of the vacuum oscillation phase as function of $y$. 
With increase of $L$ the slope decreases. 
In Fig. \ref{fig:dms-eff} the left (right) line corresponds to the short 
(long) baseline.  
  
The total oscillation phase equals
$$
\Phi = R_\Delta(y) \Phi^{\rm vac}.   
$$
Therefore at crossings of  $\pi/\left(2\Phi^{\rm vac}(y)\right)$ and  
$R_\Delta(y)$: 
\begin{equation}
\left|R_\Delta(y) \right| = \frac{\pi}{2V^{\rm vac}L}  
\label{cross}
\end{equation}
we have 
$$
\Phi (y^{\rm cross}) = \pi/2 \, \, \, \, \rightarrow 
\, \, \,  \,   
\, \, \, \sin^2 0.5\Phi (y^{\rm cross}) = 0.5.
$$
There are four crossings:  
low energy $y_l$, and $y_-$,  $y_+$ with left and right branches 
of the resonance peak as well at the resonance $y \approx 1$. 
The equation for crossing (\ref{cross}) can be written as
\begin{equation}
y^2 - 1 + r y(y - \epsilon)
= \pm \frac{\pi}{ 2V^{\rm vac}_RL} y(y^2 -1).
\label{cross2}
\end{equation}
For parts of the lines $R_\Delta(y)$, which are above the crossings 
the phase is big $\Phi > \pi/2$, while for the parts 
below the crossings the phase is small: $\Phi < \pi/2$.  

\begin{figure}
        \centering
        \includegraphics[width=\textwidth]{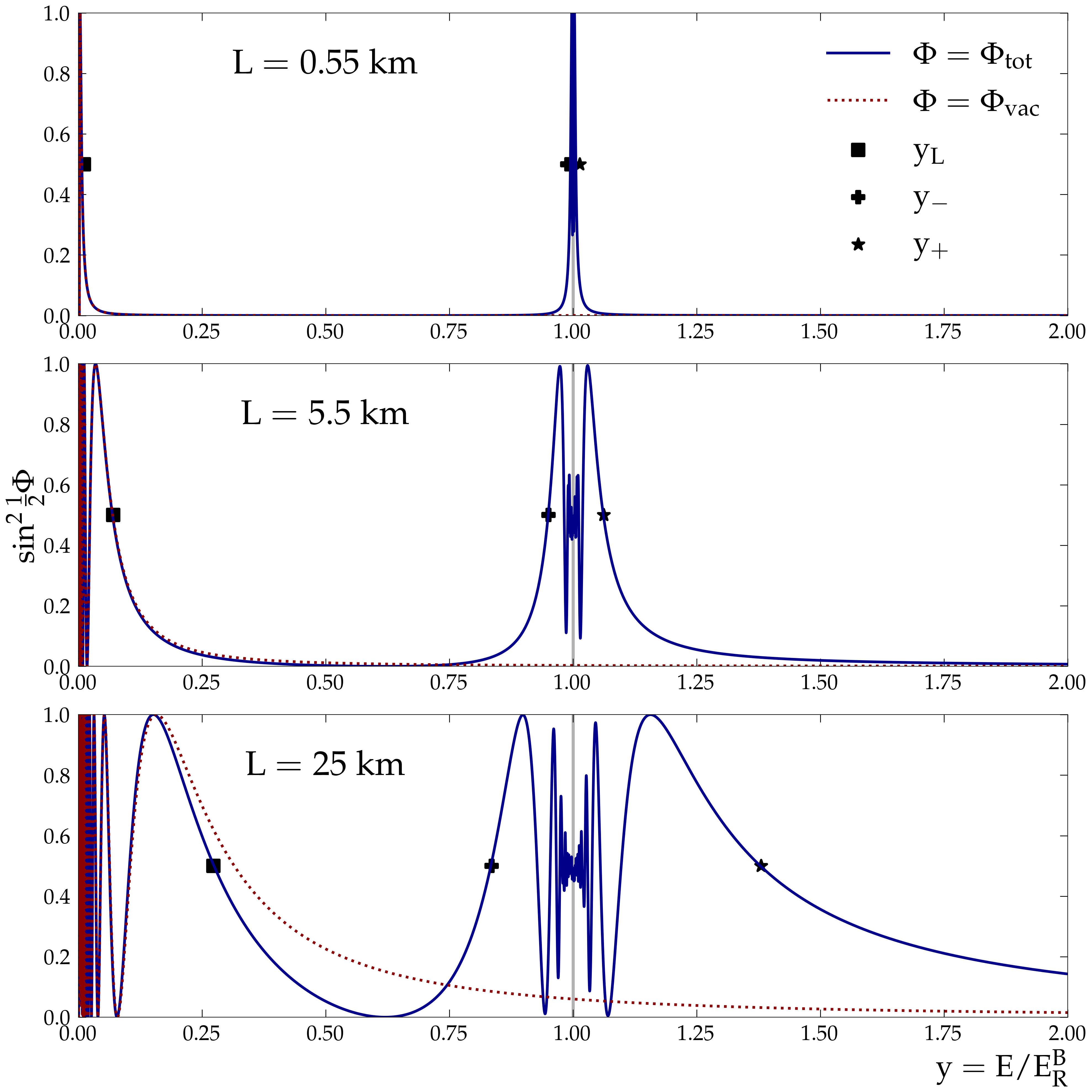}
        \caption{The  oscillatory factor as function of energy $y$  for three different 
values of baselines $L$. We take $r = 1.6$ and $\epsilon = 0$. 
The dotted lines correspond to the  oscillatory factors for pure vacuum oscillations ($r = 0$).}
        \label{fig:oscill-f}
\end{figure}

In Fig. \ref{fig:oscill-f}  we show the oscillatory factor  $\sin^2 0.5 \Phi$
as function of $y$ for three different values of baseline.   
We performed smearing over energy.

The crossings determine four intervals of $y$ with different 
observational features. 

\begin{itemize}

\item 
$y < y_l$: the oscillatory curve with  increasing period when $y \rightarrow 0$.  
At $y \rightarrow 0$  
oscillations in background  nearly coincide with the vacuum oscillations. 

\item
$y_l < y < y_-$:  oscillation dip. Here 
$\Phi < \pi/2$, the  background suppresses the phase.
 
\item
$y_- < y < y_+$: resonance interval. The phase is large: $\Phi > \pi/2$. 
In the central resonance region $\Phi \gg 1$. 

\item 
$y > y_+$: tail at high energies, $\Phi < \pi/2$,  the phase  
decreases with increase of $y$. 

\end{itemize}

With decreases  of $L$: $y_l \rightarrow 0$, while $y_-,  y_+ 
\rightarrow 1$. Thus, the dip widens, whereas  
the resonance region becomes narrower. 


\subsection{Bump: number of events}

The characteristic relevant for observations 
is not the width of the peak, but 
the energy range where the background effect is bigger than the standard 
oscillation effect.  
It is determined by the tails of resonance where $|V^B| \ll  |V^B_{\rm max}|$. 
According to  (\ref{eq:farfr})
\begin{equation}
\label{eq:resph2}
\Phi^B =  \Phi_0 \frac{y - \epsilon}{y^2 - 1},  
\end{equation}
with  
\begin{equation}
\Phi_0 \equiv V_0 L.   
\end{equation}

As a criteria  for strong effect, we can use $\sin^2 \Phi^B/2 = 1/2$, 
which gives according to 
(\ref{eq:resph2}) 
\begin{equation}
\label{eq:resph3}
y  \approx 1 \pm  \frac{\Phi_0}{\pi} (1 - \epsilon). 
\end{equation}
Therefore the region of strong effect has width  
\begin{equation}
\label{eq:resph4}
\Delta y  = \frac{2\Phi_0}{\pi} (1 - \epsilon).
\end{equation}
This region decreases with increase of $\epsilon$.

For small $\Phi_0$ the background effect  is small everywhere except for  
the resonance region. 
For instance, if $\Phi_0 = \pi/20$, then $\Delta y = 0.2$ ($\epsilon = -1$). 
In the resonance region, $E = E_R (1  \pm 0.1)$, 
we have $\sin^2 \Phi \geq 0.5$, while  
outside the resonance  $\sin^2 \Phi \approx  \sin^2 \Phi_0 = 0.024$

Let us consider total contribution from  the resonance interval.
Here the number of events  
is proportional to the integral 
\begin{equation}
\label{eq:sinresph}
I = \int^{y_{\rm max}}_{y_{\rm min}}  dy \sin^2 0.5\Phi^B(y,\epsilon) =
\int^{y_{\rm max}}_{y_{\rm min}} dy \sin^2 0.5\Phi_0 
\frac{y - \epsilon}{y^2 - 1} , 
\end{equation}
where $y_{\rm max}$  and  $y_{\rm min}$ are determined by conditions   
the phase $\Phi^B (y_{\rm max}) =   \Phi^B (y_{\rm min}) = \pi/2$, so that  
$\sin^2 0.5 \Phi^B = 1/2$.   
We  can approximate  the oscillatory factor by its 
average value:  $\sin^2 \Phi/2 \approx 0.5$.  Then 
\begin{equation}
I = 0.5 \Delta y  = \frac{\Phi_0}{\pi} (1 - \epsilon) = 
(1 - \epsilon)\frac{V_0 L}{\pi},    
\label{eq:centint}
\end{equation}
according to  (\ref{eq:resph2}), and this result is valid for $\Phi_0/\pi  \ll 1$.

More precise computation of the integral (\ref{eq:sinresph})
for any interval of $y$ can be done in the following way.
Let us introduce $\delta_y$ (which depends on
$\Phi_0$) such that in the range $|y - 1| < \delta_y$ the phase is very big:   
$\Phi_0/2 \delta_y \gg 1$,  and consequently,  the sine has 
very fast oscillations ($\delta_y \ll \Delta y$).
Then the integral $I$ can be split in three parts:
with integration over $y$ in the intervals $[1 - \delta_y \div
1 + \delta_y]$, $[0 \div 1 - \delta_y]$ and
$[1 + \delta_y \div \infty]$. 
In the first (central) interval the 
integrand can be approximated by $1/2$,  and consequently,
\begin{equation}
I = \delta_y +
\int_0^{1 -\delta_y} dy \sin^2 \frac{0.5\Phi_0 (y - \epsilon)}{y^2 - 1} +
\int_{1 + \delta_y}^{\infty} dy \sin^2 \frac{0.5\Phi_0 (y - \epsilon)}{y^2 - 1}.
\label{integration2}
\end{equation}
The tail integrals can be computed numerically as follows.  

In the central (resonance) region,
$- 2\Phi_0/\pi < (y - 1) < 2\Phi_0/\pi$, 
we can substitute the integrand $\sin^2 \Phi/2$ by $1/2$. 
Outside the resonance region, $0 < y < 1 - 2\Phi_0/\pi$
(lower region) and  $ y > 1 + 2\Phi_0/\pi$ (upper region),  
the sine squared can be approximated by 
\begin{equation}
\frac{1}{2} \left(\frac{2 \Phi_0}{\pi} \right)^2
\frac{(y - \epsilon)^2}{(y^2 - 1)^2}
\label{integrand}
\end{equation}
normalized in such a way that at the borders it equals $1/2$.
Then for small $\Phi_0/\pi$ the high and the low energy tails give 
$$
I_h \approx \frac{\Phi_0}{\pi} \frac{1 - \epsilon}{2} , \, \, \, \,
I_l \approx \frac{\Phi_0}{\pi}\frac{1 - \epsilon}{2}
\left(1 - \frac{2 \Phi_0}{\pi} \right),
$$
and the sum  equals
$$
I_{tail} = I_h + I_l \approx \frac{2 \Phi_0}{\pi}\frac{1 - \epsilon}{2}. 
$$

The ratio of the tails to the resonance (\ref{eq:centint}) contributions equals
\begin{equation}
\frac{I_{tail}}{I_c}  \approx 1 - {\cal O} \left(\frac{\Phi_0}{\pi} \right), 
\label{retiotr}
\end{equation}
and it depends on the phase weakly: with increase of
$\Phi_0$ the ratio decreases. 
The contribution from resonance width (\ref{eq:resvw}) is negligible.

\subsection{Adding usual matter effect}

The matter potential 
$V_e = \sqrt{2} G_F n_e$ does not depend on energy 
in the range we are considering.  
The equality $V_e \approx V^{\rm vac}$ determines the MSW resonance 
energy $E^{\rm MSW}$. 
Since in this setup the mixing does not change by the background,  
the MSW resonance condition has usual form:
\begin{equation}
\frac{\Delta m^2_{\rm eff}(E)}{2 E}  = \frac{V_e}{\cos 2\theta}.    
\label{y-msw}
\end{equation}

There are three possibilities depending on relative values 
of $V_e$ and $V^{\rm vac}_R$.

I. $V_e < V^{\rm vac}_R$: In this case  the refraction resonance is below 
the MSW resonance $E^B_R < E^{\rm MSW}$ (see Fig. \ref{fig:vsum-y}). 
There are three crossing of $V^{\rm vac}(y)$ with $V_e$ in the neutrino channel:  

(i) Standard MSW resonance. It  is shifted to higher energies due to 
background contribution.  
The resonance energy with background correction
can be found from eq.(\ref{y-msw}).
The expression is simplified in the case $y^{\rm MSW} \gg 1$,
so that we can take the asymptotic value
$\Delta m^2_{\rm eff}(E) \approx \Delta m^2 (1 + r)$. 
As a result,
\begin{equation}
E^{\rm MSW} = E^{\rm MSW, 0} (1 + r),
\label{eq:shift}
\end{equation}
where $E^{\rm MSW, 0}$ is the standard resonance energy 
without background:
$$
E^{\rm MSW, 0} = \cos 2\theta \frac{\Delta m^2}{2V_e}.
$$
The shift of MSW resonance can be used to search for 
the background effect. 

(ii) New crossing near refraction resonance, $y \approx 1$. 

(iii) New crossing with the low energy branch of $V^{\rm tot}$.

In the $\bar{\nu}-$ channel there are two crossings: 
(i) near the refraction resonance; 
(ii) with low energy branch of $V^{\rm sum}$. 

In the crossing points the mixing in medium
(matter plus background) becomes maximal.  

If $V_e \ll  V^{\rm vac}_R$, $E^B_R \ll E^{\rm MSW}$, 
at low energies and  short baseline experiments 
the  effects of four new crossings become 
unobservable  because in these crossings 
$\Phi \ll 1$. 

II.  $V_e > V^{\rm vac}_R$: In this case 
the refraction resonance is above the MSW resonance:  $E^B_R > E^{\rm MSW}$. 
Depending on $\epsilon$ the shift of  the MSW resonance can  be to 
higher or low energies. 

As before there are two new crossings in the $\nu-$channel 
and two new crossings in the $\bar{\nu}-$channel. 
In the $\nu-$channel one crossing is near the refraction resonance, 
and another one is in the high energy branch. The energy of the latter can be 
substantially larger than $y = 1$.  In the $\bar{\nu}-$channel the two crossings are near 
the refraction resonance being  in the low energy branch.

III. The case  of $V_e \approx V^{\rm vac}_R$ is of special interest: 
the  standard MSW resonance coincides with the refraction resonance, 
while two new resonances (at $y > 1$ and  at $y < 1$) can be far from 
the refraction resonance $y = 1$.

\subsection{Generation of mixing in the background}

In the previous consideration the matrix of
potentials had only one entry and so it was diagonal in the mass eigenstate basis.
If  couplings of other mass states with background are not neglected
the transition $\nu_1 \bar{\chi} \rightarrow \nu_2 \bar{\chi}$
generates a non-diagonal element  of the matrix of potentials which is proportional
$g_1 g_2^*$. In the $2\nu$ case the total Hamiltonian becomes  
\begin{equation}
\label{eq:hami}
H^B = 
\left(\begin{matrix}
0 & \alpha V^B\\
\alpha^* V^B & V^{\rm vac} + V^B  
\end{matrix}\right)  
=  V^{\rm vac} \left(\begin{matrix}
0 & \alpha \kappa \\
\alpha^* \kappa & 1 + \kappa  
\end{matrix}
\right),
\end{equation}
where 
$$
\alpha \equiv \frac{g_1 g_2^*}{|g_{2}|^2 - |g_{1}|^2},  
$$
$V^B = V^B (|g|^2 \rightarrow |g_{2}|^2 - |g_{1}|^2)$   
and $V^B (|g|^2)$  is the background potential discussed 
in the previous sections. $\kappa$ is defined in  
(\ref{ratio-v}). 

Notice that the resonance energies are  different for
different neutrino mass states $\nu_i$:
$$
E_{2R} - E_{1R} = \frac{\Delta m^2}{2m_\chi} \ll E_R, 
$$
but this difference is still much smaller than the scale of smearing due to
motion of scatterers. 
Therefore we can neglect dependence of potentials on the
neutrino masses  and the only relevant dependence on type of neutrino
is in the coupling constants. 

Diagonalization of the matrix (\ref{eq:hami}) gives the difference of eigenvalues  
\begin{equation}
R_\Delta  = 
\sqrt{\left(1 + \kappa \right)^2  + \left(2 \alpha \kappa \right)^2}, 
\label{eq:effdms}
\end{equation}
and the mixing angle of mass states 
\begin{equation}
\sin^2 2 \theta^B = \frac{\left(2 \alpha \kappa \right)^2 }  
{\left(1 + \kappa \right)^2  + \left(2 \alpha \kappa \right)^2}. 
\label{eq:effmix}
\end{equation}
The flavor mixing angle becomes 
\begin{equation}
\label{eq:effdms1}
\theta_f = \theta + \theta^B. 
\end{equation}

Let us consider different limits and benchmark points. 

1. $y \rightarrow 0$: $R_\Delta \rightarrow 1$ and  $\theta^B \rightarrow 0$. 
The background effect is negligible.

2. $y \rightarrow y_c$: the cancellation point ($V^B = - V^{\rm vac}$) 
becomes the energy of MSW resonance on the background. Here  
the mixing is maximal $\sin^2 2\theta^B = 1$ and 
splitting is non-zero: 
$$
R_\Delta  = 2 \alpha. 
$$
The transition probability equals
$P \approx \sin^2 (\alpha \Phi^{\rm vac})$. 

3. In the peak, $y \approx 1$: $V^B \gg V^{\rm vac}$,  the ratio equals 
$$
R_\Delta = \frac{V^B}{V^{\rm vac}} \sqrt{1 + 4\alpha^2} = \kappa \sqrt{1 + 4\alpha^2},  
$$
and the angle is 
$$
\sin^2 2\theta^B =  \frac{4\alpha^2 }{1 + 4\alpha^2 }. 
$$

4. In the refraction resonance, $y = 1$ ($V^B = 0$): $R_\Delta = 1$ and $\theta^B = 0$. 

5. In asymptotics $y \rightarrow \infty$: $V^B/ V^{\rm vac} \rightarrow r$. Correspondingly, 
$$
R_\Delta  = \sqrt{\left( 1 + r\right)^2 + \left(2\alpha r\right)^2 }, 
$$
and 
$$
\sin^2 2\theta^B =  
\frac{\left(2\alpha r\right)^2 }{\left(1 + r \right)^2 + \left(2\alpha r\right)^2 }. 
$$

The transition probability  equals 
$$
P = \sin^2 2(\theta + \theta^B) \sin^2 (\Phi^{\rm vac} R/2). 
$$

For small $\alpha$ in comparison to no-mixing case modifications of $P$ are small. 
The most significant change is in the cancellation region.  

Finally,  let us comment on the case of three different fermions
$\chi_j$ - each per generation. If VEV of $\phi$ is the only source of
neutrino mass then the couplings are diagonal in the mass basis.
Furthermore, transition $\nu_i \bar{\chi_i} \rightarrow
\nu_j \bar{\chi_j}$ will not form potential,  since
final  $\bar{\chi_j}$ differs  from initial $\bar{\chi_i}$
being orthogonal each other. In this case the matrix of
potentials is diagonal and the difference of  diagonal
elements, $V = V_i - V_j \propto |g_i^2| - |g_j^2|$, 
will enter expressions considered above.

\subsection{Resonance refraction and  Glashow resonance}

In the standard model, the resonance refraction
is realized in the Glashow resonance:
that is, in the  $\nu_e - e$ scattering with $W$ boson
as the mediator. The resonance energy equals 
$E_R = m_W^2/2m_e \approx 6.4$ PeV. 
Dependence of the potential on neutrino energy
is described by the Eq. (\ref{Eq:poten-dim}) with
$\epsilon = -1$,
$V_0 = \sqrt{2} G_F n_e$ and
$\xi = 3 g^2_W/16 \pi$. 

At low energies the potential coincides with the
Wolfenstein potential.
The difference from what we have discussed before is that
the coupling is large $g_W \sim {\cal O}(1)$. Therefore
the width of the resonance is not negligible, 
enhancement is not extremely strong and smearing effect is weaker. 
The  maxima
$$
|V_{\rm max}|  = V_0 \frac{m_W}{\Gamma_W} = V_0 \frac{16 \pi}{3 g_W^2}
$$
are achieved at $E = E_R (1 \pm \Gamma_W/m_W)$. 
In the resonance region the vacuum contribution, $\Delta m^2/2E$ 
is negligible: $r = 10^{-6}$. Vacuum mixing is strongly suppressed.
Furthermore, dumping due to absorption can be substantial. 

The refraction length in resonance can be reduced by factor
20 in comparison to the Wolfenstein length,  being of the order 300 km. 
However, existence  of  observable effects
at the Earth is questionable.

1. Oscillation effects with usual $\Delta m^2$ and $\theta$ 
are negligible.  Refraction index is still very close to 1, so that
bending and refraction effects are negligible too.

2. One can explore  possible effect in astrophysical
objects - sources of high energy neutrinos.

3. Mixing of active neutrinos with sterile neutrinos of mass
$10^2$ eV can be considered. In this case $\Delta m^2/ 2 E_R \sim V_e$ and
the mixing can be  enhanced in matter.

\section{Applications to specific experiments}

\subsection{Signatures and implications}

Recall that the oscillatory pattern in terms of universal variables, 
$R_\Delta(y)$ and $y$ depends on 
(i) $r$ - relative strength of interactions with background (\ref{ratio-vres}), 
(ii) $\epsilon$ - charge asymmetry of the background, 
(iii) baseline $L$. 
Thus, observing the oscillatory pattern at given $L$ one can determine $\epsilon$  and 
$r$ (which is the combination of the 
fundamental parameters and density of a background (\ref{ratio-vres}))  
or put bounds on these parameters. 


Observable effects of the  background 
vanish completely if $r \rightarrow 0$, 
however,  they do not disappear when $\epsilon \rightarrow 1$.  
At $\epsilon = 1$ the resonance is absent, the  cancellation point  
is at $y_c = 1$ and 
\begin{equation}
R_\Delta(y)  = 1 +  r \, \frac{y}{y + 1}. 
\label{eps1}
\end{equation}
So, the corrections increase with $y$: at $y = 1$ the ratio equals 
$R_\Delta = 1 + r/2$ , 
for $y \rightarrow \infty$: $R_\Delta  =  1 + r$. For large energies 
the background effects are determined by $r$ 
and dependence on $\epsilon$ is weak.

To some extend the effects of $\epsilon$ and $r$ on the oscillatory pattern 
correlate, and there is certain degeneracy. However, variations of the pattern with 
$r$ can be much more substantial  than that with 
$\epsilon$. Effect of $\epsilon$ is restricted by its minimal value $-1$.

The presence of the resonance bump testifies for $\epsilon \neq 1$. 
Value of $\epsilon$ determines the benchmark energies. 
With $\epsilon \rightarrow -1$ the region of distortions in the resonance 
interval becomes wider. 
Measuring the oscillatory pattern in different energy ranges allows to disentangle 
effects of $r$ and  $\epsilon$.  
Let us summarize signatures of interactions with background. For $\kappa > 0$ they include: 

\begin{itemize}

\item 
deviation of  the oscillatory pattern from $\sin^2 (A/y)$  
in the low energy interval;  

\item 
oscillation dip at $y < 1$, with zero at $y_c$; 

\item 
increase of the probability at $y \rightarrow 1$; 

\item
 bump at $y \sim 1$; 

\item 
tail at $y > 1.2$,  which corresponds to larger $\Delta m^2_{\rm eff}$ than at low energies. 

\end{itemize}

In the presence of usual matter we have in addition 
\begin{itemize}

\item
shift of the MSW resonance to larger (if $E^{\rm MSW} > E_R^B$ ) or smaller (if $E^{\rm MSW} <  E_R^B$ ) 
energies;

\item 
appearance of new MSW resonances. 

\end{itemize}

For $\kappa < 0 $ the dip is at higher energies. In asymptotics the effective  
$\Delta m^2_{\rm eff}$ is smaller than that at low energies. 

Thus, to search for effects 
for fixed $L$ one can consider different energy intervals. 
For a given neutrino beam one can use different $L$, e.g.,  
results from near and far  detectors.

\subsection{MiniBooNE excess and resonance refraction}

The low energy excess of events reported by the MiniBooNE collaboration 
\cite{collaboration2020updated} could be a manifestation 
of the resonance refraction  \citep{Asaadi_2018}. The background is composed of the overdense  
relic neutrinos. 
In this case $m_\chi = 0.05$ eV and $\epsilon \approx 0$.

The best fit of the MiniBooNE data is obtained for values of parameters  
\begin{equation}
E^B_R = (320 - 340) \,  {\rm MeV}, \,  \, \, \, \,  \,  \,   
Y \equiv  \frac{g^2 (n_\chi + \bar{n}_\chi)}{8 m_\chi} \geq 10^{-3} \,  {\rm eV}^2.  
\label{bf-point}
\end{equation}
Then 
\begin{equation}
m_\phi = \sqrt{2 m_\chi E^B_R} = 5.8\,  {\rm keV}. 
\label{mphi2}
\end{equation}
Notice that the mediator and background  particles  are light enough and therefore the 
astrophysical bounds on $g$ are applicable (see sect. 2.1). 

From (\ref{bf-point})  we obtain 
\begin{equation}
V_0^B = \frac{2Y}{E_R^B} =  5.9 \cdot 10^{-12} \, {\rm eV}, ~~~~
V_R^{\rm vac}  =  3.7 \cdot 10^{-12} \, {\rm eV}, 
\label{vacpot}
\end{equation}
and correspondingly,  
$$
r =  \frac{4Y}{\Delta m^2} = 1.59, \, \, \, \, \,  y_c = 0.62.  
$$
Thus, in the resonance region and above it the 
background potential dominates. 
The usual matter potential is very small: 
$V_e = 2 \cdot 10^{-13} \, {\rm eV}$.

The MiniBooNE baseline $L_{\rm MB} = 541$ m corresponds to 
$$
\frac{1}{L_{\rm MB}} = 3.1 \cdot 10^{-10} \, {\rm eV} \gg 
V_R^{\rm vac}, \, V_0^B,
$$
which means that the phase is very small, $\Phi \ll 1$,  everywhere except for narrow region 
close to $y = 1$. The resonance peak is smeared by the energy resolution. 

\begin{figure}
        \centering
        \includegraphics[width=\textwidth]{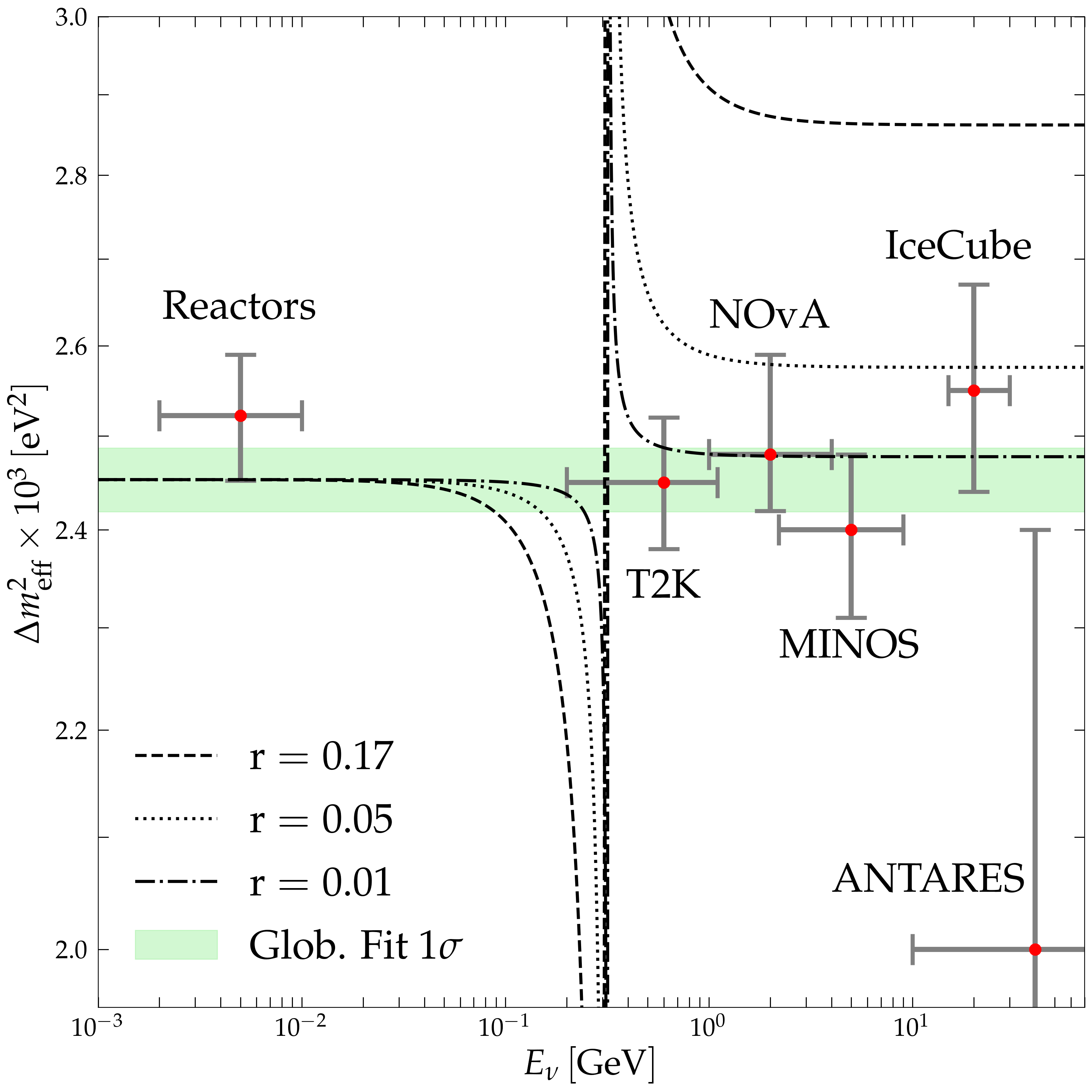}
        \caption{The effective mass squared difference as function of the neutrino 
energy for different values or $r$. 
The curves are normalized at $E \rightarrow  0 $ to the value of the $\Delta m_{32}^2$ 
from the global fit of all the data.  Explanation of the MiniBooNE  requires 
$r > 1.6$. Shown are the values of $\Delta m_{32}^2$, $\Delta m_{31}^2$  and 
$\Delta m_{ee}^2$ extracted from experiments at different energies. }
        \label{fig:meff-data}
\end{figure}

Let us show that this solution is excluded because of strong 
dependence of the  effective $\Delta m^2_{\rm eff}$ on energy ($y$).  
In Fig. \ref{fig:meff-data} we show $\Delta m^2_{\rm eff}$ as function of energy 
for $E_R = 320$ MeV, $\epsilon = 0$
and different values of $r$. 
At low energies $y \ll 1$, 
$\Delta m^2_{\rm eff} \approx \Delta m^2$ (as in vacuum) while  above the resonance 
\begin{equation}
\Delta m^2_{\rm eff}  \approx \left(1 +  r \frac{y^2}{y^2 -1}\right)\Delta m^2.
\label{vacpot}
\end{equation}
According to this equation 
for $y = 2$ and $y = 3$, which correspond to $E = 680$ MeV and $1020$ MeV,  
the enhancement of $\Delta m^2_{\rm eff}$  is given by factors 
$3.12$ and  $2.79$ respectively. 
In asymptotics, $y \rightarrow \infty$, it converges to 2.59. 

Fig. \ref{fig:meff-data} shows also results of measurements of 
the ``atmospheric" $\Delta m^2 \approx m_3^2$  
at different energies. 
At the  lowest  energies, $E = (2 - 5)$ MeV ($y \sim 10^{-2}$)  
the data on $\Delta m^2_{ee} \approx \Delta m^2_{31}$ are provided by the reactor experiments 
\cite{Adey:2018zwh,Bak:2018ydk,DoubleChooz:2019qbj}. Here the background effect can be neglected. 
The T2K experiment \cite{Abe:2020vot,Abe:2019vii} measures $\Delta m^2_{32}$  at $(0.3 - 1.3)$ GeV  
which is slightly above the resonance. At higher energies (essentially in asymptotics) the data 
are given by NOvA \cite{Acero:2019ksn} and then MINOS and MINOS$+$ \cite{Adamson:2020ypy}.  
At even higher energies 
IceCube DeepCore \cite{Aartsen:2019tjl} and ANTARES \cite{Albert:2018mnz} give information on  $\Delta m^2_{32}$. 

The main conclusion is that within the experimental error bars $\Delta m^2_{\rm eff}$
does not depend on energy over 4 orders of magnitude. This puts strong bound 
on strength of interaction with background: 
\begin{equation}
r \lesssim 0.01, 
\label{eq:boundr}
\end{equation}
which certainly excludes $r >  1.6$ required by MiniBooNE explanation.

Similar result can be obtained for negative $r$. Now above the resonance 
the predicted values of $\Delta m^2_{\rm eff}$ are  below the experimental points. 

The same consideration with the same conclusion is applied for the bosonic background and 
fermionic mediator. 
In particular, the  Fig. \ref{fig:meff-data} will be unchanged. 
The only difference  is that the potential is 2 times larger which can be accounted by 
renormalization $g \rightarrow \sqrt{2} g$. The latter could have some implications to 
particle physics model but not to the exclusion.

\subsection{Bounds on the background effects}

We have obtained the upper bound
on strength of the background effects $r$ (\ref{eq:boundr})
for $E_R \sim 320$ MeV.
According to Fig. \ref{fig:meff-data} similar bound can be established in the interval
of $E_R$ from  $10$ MeV to  $10$ GeV. For $E_R < 1$ MeV - no distortion
is expected in the region of observations ({\it i.e.} at $E > $ MeV),  while for
$E_R > 10$ GeV the effect of background in the observable region becomes much
smaller than vacuum effect and it decreases with energy decrease.

The strength $r$  (\ref{ratio-vres}) can be written as 
\begin{equation}
r  = \frac{2V_0 E_R}{\Delta m^2}.
\label{eq:strength}
\end{equation}
This means that for given $E_R$ and $r$ the
potential is restricted by
\begin{equation}
V_0 = r \frac{\Delta m^2}{2E_R}.
\label{eq:strength}
\end{equation}
The largest value of $E_R$, for which a given bound on $r$
exists,  gives the most strong restriction  on $V_0$.
Therefore according to (\ref{eq:strength})
\begin{equation}
r (E = 0.32 ~{\rm GeV}) = r(E = 3~ {\rm GeV}) \frac{0.32~ {\rm GeV}}{3~ {\rm GeV}}
\approx 10^{-3}.
\label{eq:strength2}
\end{equation}
Thus, consideration at higher energies allows to
strengthen the bound on $r$.

For the background particles at rest
the strength factor can be written as
\begin{equation}
r = \frac{g^2 n_\chi}{2 m_\chi \Delta m^2}.
\label{eq:strength2}
\end{equation}

Is the bound on $r$ we obtained  from resonance
refraction substantial, or there are other more strong bounds?
One such a bound on the system comes from contribution
of $\chi$ to the dark matter in the Universe:
\begin{equation}
\rho_\chi = E_\chi n_\chi \geq m_\chi n_\chi.
\label{eq:rho1}
\end{equation}
For a given value of $m_\chi$ this gives the number density
of $\chi$ which compose $\rho_\chi/ \rho_{\rm DM}$ fraction of the
local dark matter:
\begin{equation}
n_\chi \sim \frac{\rho_{\rm DM}}{m_\chi}
\frac{\rho_\chi}{\rho_{\rm DM}}.
\label{eq:nchi}
\end{equation}
Inserting this expression into
(\ref{eq:strength2}) and taking
for the local energy density of DM $\rho_{\rm DM} = 0.4~{\rm GeV}/{\rm cm}^3$,
we obtain the strength factor
\begin{equation}
r = 2.6 \cdot 10^{-7}
\left(\frac{g^2}{10^{-3}}\right)^2
\left(\frac{0.05 {\rm eV}}{m_\chi}\right)^2
\left(\frac{\rho_\chi}{\rho_{\rm DM}}\right).
\label{eq:strength3}
\end{equation}

For $g$ satisfying the bound (\ref{eq:gbound}), $\rho_\chi = \rho_{\rm DM}$
and $m_\chi = 0.05$ eV Eq. (\ref{eq:strength3}) gives 
$r = 2.6 \cdot 10^{-14}$ which is much below the refraction bound.
For these values of parameters $n_\chi = 8 \cdot 10^9$ cm$^{-3}$.
$r$ can be enhanced if we take
smaller mass of  $\chi$ and $g = 10^{-3}$,  which satisfies
the laboratory bounds but  requires
more complicated cosmological evolution that  allows to avoid
BBN and CMB bounds. Then $r = 10^{-3}$ can be obtained for
$m_\chi = 8 \cdot 10^{-4}$ eV.
The corresponding number density of $\chi$ equals
$n_\chi = 5 \cdot 10^{11}$ cm$^{-3}$.

This consideration is valid for bosonic background with  changing subscripts 
$\chi \leftrightarrow \phi$ in Eq. (\ref{eq:nchi} - \ref{eq:strength3}). 
For the fermionic background additional restrictions
follow from Pauli principle.
Indeed, the density indicated above gives
the Fermi momentum of the degenerate gas
$p_F = (6 \pi^2 n_\chi)^{1/3} = 1.3$ eV.
That is, $E_\chi \approx p_F \gg m_\chi$,
and therefore we deal here with strongly degenerate
fermion gas. Consequently,  in all considerations above
we should substitute 
$$
m_\chi \rightarrow E_\chi \sim p_F = (6 \pi^2 n_\chi)^{1/3}.
$$ 
In particular,
$E_R = m_\phi^2/2 E_\chi$ and
\begin{equation}
r = \frac{g^2}{2\Delta m^2}
\frac{n_\chi}{E_\chi} =
\frac{g^2}{2(6 \pi^2)^{1/3})\Delta m^2} (n_\chi)^{2/3}. 
\label{eq:rdegf}
\end{equation}
Using expression for the energy density in $\chi$
\begin{equation}
\rho_\chi = E_\chi n_\chi =
(6 \pi^2)^{1/3} n_\chi^{4/3}
\label{eq:rhochi}
\end{equation}
we obtain
\begin{equation}
r = \frac{g^2 \sqrt{\rho_\chi}}{2\sqrt{6} \pi \Delta m^2}.
\label{eq:rdegf2}
\end{equation}
Numerically this gives
\begin{equation}
r = 4.7 \cdot 10^{-8}
\left(\frac{g^2}{10^{-3}}\right)^2
\sqrt{\frac{\rho_\chi}{\rho_{\rm DM}}}.
\label{eq:rdegf3}
\end{equation}
Thus, $r$  is determined by the coupling constant and
fraction of the DM in  $\chi$ and does not depend on $m_\chi$.
The value  $r \leq 4.7 \cdot 10^{-8}$,  which is much smaller than
sensitivity range to the resonance refraction effects of experiments at the laboratory
energies.

\section{Conclusions} 

1. In general, the  medium potential is function of the neutrino energy
and this function  depends on the C-asymmetry of a background. 
The energy dependence of $V^B$ may have a resonance character related to
the exchange of (on shell) mediator
of interactions.  Resonance is realized at $\sqrt{s} = M_{med}$ and 
for light mediators and light scatterers
(which requires extension of the Standard model) the resonance refraction can occur at
energies available at laboratories.

2. The relative correction to the vacuum (kinetic)
term from background vanishes at low energies, it can dominate
in resonance and above it. At high energies the correction converges to
constant. The interplay of the energy dependent potential
$V^B(y)$ and vacuum contribution $V^{\rm vac}(y)$ has several
important features:

- Cancellation of the contributions which corresponds
to the MSW resonance on background (when mixing in the background
is introduced),

- above the resonance $V^B(y)$ gives correction to
$V^{\rm vac}(y)$ which does not disappear in asymptotics
$E \rightarrow \infty$.

3. The background can produce mixing of mass states, 
that is,  the non-diagonal matrix of potentials in the
mass basis. For small mixing substantial effect on oscillations
appears in the region around the cancellation point (the MSW resonance on a background).

4. For long-baseline experiments usual matter effect should
be added. The interaction with background   shifts the energy of MSW resonance
(which provides important signature)   and leads to  appearance of new resonances
around $E_R^B$.

5. Signatures of refraction on the background include:
(i) deviation of the oscillatory pattern in energy
from $\sin^2 (A/E)$, (ii)  dip of the  oscillation probability
below or above resonance, (iii) bump in the resonance region,
(iv) additional contribution to $V^{\rm vac}(y)$ above
refraction resonance  which does not disappear in asymptotics.

6. Effects of background can be considered
as  modification of the effective $\Delta m^2_{\rm eff} (y)$  with peculiar
dependence on energy.

7. As an  example we applied our results to the 
MiniBooNE excess interpreted as bump produced by the refraction resonance. 
We show that this interpretation is excluded  because of strong  difference of 
$\Delta m^2_{\rm eff}$ expected at high energies  
(T2K, NOvA, MINOS, MINOS+, IceCube, ANTARES)  and low energies 
(reactor experiments) in contrast to observations.  We obtain the bound on 
the relative strength of neutrino interactions with background $r < (0.001 - 0.01)$.

\section*{Acknowledgements}

A.Y.S.  thanks E. Kh. Akhmedov   for  useful discussions.
V.B.V. is grateful for support from the {\sc Villum Fonden} through the project no.~29388, 
and the ICTP Postgraduate Diploma Programme.

\printbibliography

@article{Wolfenstein-PhysRevD,
  title = {Neutrino oscillations in matter},
  author = {Wolfenstein, L.},
  journal = {Phys. Rev. D},
  volume = {17},
  issue = {9},
  pages = {2369--2374},
  numpages = {0},
  year = {1978},
  month = {5},
  publisher = {American Physical Society},
  doi = {10.1103/PhysRevD.17.2369},
  url = {https://link.aps.org/doi/10.1103/PhysRevD.17.2369}
}

@article{Opher:1974drq,
	author = "Opher, R.",
	title = "{Coherent scattering of cosmic neutrinos}",
	journal = "Astron. Astrophys.",
	volume = "37",
	number = "1",
	pages = "135--137",
	year = "1974"
}

@article{Barger:1980tf,
	author = "Barger, Vernon D. and Whisnant, K. and Pakvasa, S. and Phillips, R. J. N.",
	title = "{Matter Effects on Three-Neutrino Oscillations}",
	reportNumber = "DOE-ER/00881-152",
	doi = "10.1103/PhysRevD.22.2718",
	journal = "Phys. Rev. D",
	volume = "22",
	pages = "2718",
	year = "1980"
}

@article{Langacker:1982ih,
	author = "Langacker, Paul and Leveille, Jacques P. and Sheiman, Jon",
	title = "{On the Detection of Cosmological Neutrinos by Coherent Scattering}",
	reportNumber = "UM HE 82-28",
	doi = "10.1103/PhysRevD.27.1228",
	journal = "Phys. Rev. D",
	volume = "27",
	pages = "1228",
	year = "1983"
}

@article{LUNARDINI2000260,
	title = {The minimum width condition for neutrino conversion in matter},
	journal = {Nuclear Physics B},
	volume = {583},
	number = {1},
	pages = {260-290},
	year = {2000},
	issn = {0550-3213},
	doi = {https://doi.org/10.1016/S0550-3213(00)00341-2},
	url = {https://www.sciencedirect.com/science/article/pii/S0550321300003412},
	author = {C. Lunardini and A.Yu. Smirnov}
}

@article{Asaadi_2018,
	title={New light Higgs boson and short-baseline neutrino anomalies},
	volume={97},
	ISSN={2470-0029},
	url={http://dx.doi.org/10.1103/PhysRevD.97.075021},
	DOI={10.1103/physrevd.97.075021},
	number={7},
	journal={Physical Review D},
	publisher={American Physical Society (APS)},
	author={Asaadi, J. and Church, E. and Guenette, R. and Jones, B. J. P. and Szelc, A. M.},
	year={2018},
	month={4}
}

@misc{collaboration2020updated,
	title={Updated MiniBooNE Neutrino Oscillation Results with Increased Data and New Background Studies},
	author={MiniBooNE Collaboration and A. A. Aguilar-Arevalo and B. C. Brown and J. M. Conrad and R. Dharmapalan and A. Diaz and Z. Djurcic and D. A. Finley and R. Ford and G. T. Garvey and S. Gollapinni and A. Hourlier and E. -C. Huang and N. W. Kamp and G. Karagiorgi and T. Katori and T. Kobilarcik and K. Lin and W. C. Louis and C. Mariani and W. Marsh and G. B. Mills and J. Mirabal-Martinez and C. D. Moore and R. H. Nelson and J. Nowak and I. Parmaksiz and Z. Pavlovic and H. Ray and B. P. Roe and A. D. Russell and M. H. Shaevitz and J. Spitz and I. Stancu and R. Tayloe and R. T. Thornton and M. Tzanov and R. G. Van de Water and D. H. White and E. D. Zimmerman},
	year={2020},
	eprint={2006.16883},
	archivePrefix={arXiv},
	primaryClass={hep-ex}
}

@article{Nieves:2018vxl,
	author = "Nieves, Jos\'e F. and Sahu, Sarira",
	title = "{Neutrino effective potential in a fermion and scalar background}",
	eprint = "1808.01629",
	archivePrefix = "arXiv",
	primaryClass = "hep-ph",
	doi = "10.1103/PhysRevD.98.063003",
	journal = "Phys. Rev. D",
	volume = "98",
	number = "6",
	pages = "063003",
	year = "2018"
}

@article{Nieves:2018ewk,
	author = "Nieves, Jos\'e F. and Sahu, Sarira",
	title = "{Neutrino damping in a fermion and scalar background}",
	eprint = "1812.05672",
	archivePrefix = "arXiv",
	primaryClass = "hep-ph",
	doi = "10.1103/PhysRevD.99.095013",
	journal = "Phys. Rev. D",
	volume = "99",
	number = "9",
	pages = "095013",
	year = "2019"
}

@article{Ge:2018uhz,
	author         = "Ge, Shao-Feng and Parke, Stephen J.",
	title          = "{Scalar Nonstandard Interactions in Neutrino
	Oscillation}",
	journal        = "Phys. Rev. Lett.",
	volume         = "122",
	year           = "2019",
	number         = "21",
	pages          = "211801",
	doi            = "10.1103/PhysRevLett.122.211801",
	eprint         = "1812.08376",
	archivePrefix  = "arXiv",
	primaryClass   = "hep-ph",
	reportNumber   = "IPMU18-0206, FERMILAB-PUB-18-487-T",
	SLACcitation   = "%%CITATION = ARXIV:1812.08376;%%"
}

@article{Choi:2019zxy,
	author = "Choi, Ki-Young and Chun, Eung Jin and Kim, Jongkuk",
	title = "{Neutrino Oscillations in Dark Matter}",
	eprint = "1909.10478",
	archivePrefix = "arXiv",
	primaryClass = "hep-ph",
	doi = "10.1016/j.dark.2020.100606",
	journal = "Phys. Dark Univ.",
	volume = "30",
	pages = "100606",
	year = "2020"
}

@article{Babu:2019iml,
	author = "Babu, K. S. and Chauhan, Garv and Bhupal Dev, P. S.",
	title = "{Neutrino nonstandard interactions via light scalars in the Earth, Sun, supernovae, and the early Universe}",
	eprint = "1912.13488",
	archivePrefix = "arXiv",
	primaryClass = "hep-ph",
	reportNumber = "OSU-HEP-19-11",
	doi = "10.1103/PhysRevD.101.095029",
	journal = "Phys. Rev. D",
	volume = "101",
	number = "9",
	pages = "095029",
	year = "2020"
}

@article{Choi:2020ydp,
	author = "Choi, Ki-Young and Chun, Eung Jin and Kim, Jongkuk",
	title = "{Dispersion of neutrinos in a medium}",
	eprint = "2012.09474",
	archivePrefix = "arXiv",
	primaryClass = "hep-ph",
	reportNumber = "KIAS-P20072",
	month = "12",
	year = "2020"
}

@article{Ge:2019tdi,
	author = "Ge, Shao-Feng and Murayama, Hitoshi",
	title = "{Apparent CPT Violation in Neutrino Oscillation from Dark Non-Standard Interactions}",
	eprint = "1904.02518",
	archivePrefix = "arXiv",
	primaryClass = "hep-ph",
	reportNumber = "IPMU19-0047",
	month = "4",
	year = "2019"
}

@article{Berlin:2016woy,
	author         = "Berlin, Asher",
	title          = "{Neutrino Oscillations as a Probe of Light Scalar Dark
	Matter}",
	journal        = "Phys. Rev. Lett.",
	volume         = "117",
	year           = "2016",
	number         = "23",
	pages          = "231801",
	doi            = "10.1103/PhysRevLett.117.231801",
	eprint         = "1608.01307",
	archivePrefix  = "arXiv",
	primaryClass   = "hep-ph",
	SLACcitation   = "%%CITATION = ARXIV:1608.01307;%%"
}

@article{Rodejohann:2017vup,
	author         = "Rodejohann, Werner and Xu, Xun-Jie and Yaguna, Carlos E.",
	title          = "{Distinguishing between Dirac and Majorana neutrinos in
	the presence of general interactions}",
	journal        = "JHEP",
	volume         = "05",
	year           = "2017",
	pages          = "024",
	doi            = "10.1007/JHEP05(2017)024",
	eprint         = "1702.05721",
	archivePrefix  = "arXiv",
	primaryClass   = "hep-ph",
	SLACcitation   = "%%CITATION = ARXIV:1702.05721;%%"
}

@article{Lindner:2018kjo,
	author         = "Lindner, Manfred and Queiroz, Farinaldo S. and
	Rodejohann, Werner and Xu, Xun-Jie",
	title          = "{Neutrino-electron scattering: general constraints on
	$Z'$ and dark photon models}",
	journal        = "JHEP",
	volume         = "05",
	year           = "2018",
	pages          = "098",
	doi            = "10.1007/JHEP05(2018)098",
	eprint         = "1803.00060",
	archivePrefix  = "arXiv",
	primaryClass   = "hep-ph",
	SLACcitation   = "%%CITATION = ARXIV:1803.00060;%%"
}

@article{Arcadi:2019uif,
	author         = "Arcadi, Giorgio and Lindner, Manfred and Martins, Jessica
	and Queiroz, Farinaldo S.",
	title          = "{New Physics Probes: Atomic Parity Violation, Polarized
	Electron Scattering and Neutrino-Nucleus Coherent
	Scattering}",
	year           = "2019",
	eprint         = "1906.04755",
	archivePrefix  = "arXiv",
	primaryClass   = "hep-ph",
	SLACcitation   = "%%CITATION = ARXIV:1906.04755;%%"
}

@article{Lindner:2016wff,
	author         = "Lindner, Manfred and Rodejohann, Werner and Xu, Xun-Jie",
	title          = "{Coherent Neutrino-Nucleus Scattering and new Neutrino
	Interactions}",
	journal        = "JHEP",
	volume         = "03",
	year           = "2017",
	pages          = "097",
	doi            = "10.1007/JHEP03(2017)097",
	eprint         = "1612.04150",
	archivePrefix  = "arXiv",
	primaryClass   = "hep-ph",
	SLACcitation   = "%%CITATION = ARXIV:1612.04150;%%"
}

@article{Farzan:2018gtr,
	author         = "Farzan, Yasaman and Lindner, Manfred and Rodejohann,
	Werner and Xu, Xun-Jie",
	title          = "{Probing neutrino coupling to a light scalar with
	coherent neutrino scattering}",
	journal        = "JHEP",
	volume         = "05",
	year           = "2018",
	pages          = "066",
	doi            = "10.1007/JHEP05(2018)066",
	eprint         = "1802.05171",
	archivePrefix  = "arXiv",
	primaryClass   = "hep-ph",
	SLACcitation   = "%%CITATION = ARXIV:1802.05171;%%"
}

@article{Brdar:2018qqj,
	author         = "Brdar, Vedran and Rodejohann, Werner and Xu, Xun-Jie",
	title          = "{Producing a new Fermion in Coherent Elastic
	Neutrino-Nucleus Scattering: from Neutrino Mass to Dark
	Matter}",
	journal        = "JHEP",
	volume         = "12",
	year           = "2018",
	pages          = "024",
	doi            = "10.1007/JHEP12(2018)024",
	eprint         = "1810.03626",
	archivePrefix  = "arXiv",
	primaryClass   = "hep-ph",
	SLACcitation   = "%%CITATION = ARXIV:1810.03626;%%"
}

@article{Bjorken:2009mm,
	author         = "Bjorken, James D. and Essig, Rouven and Schuster, Philip
	and Toro, Natalia",
	title          = "{New Fixed-Target Experiments to Search for Dark Gauge
	Forces}",
	journal        = "Phys. Rev.",
	volume         = "D80",
	year           = "2009",
	pages          = "075018",
	doi            = "10.1103/PhysRevD.80.075018",
	eprint         = "0906.0580",
	archivePrefix  = "arXiv",
	primaryClass   = "hep-ph",
	reportNumber   = "SLAC-PUB-13650, SU-ITP-09-22",
	SLACcitation   = "%%CITATION = ARXIV:0906.0580;%%"
}

@article{Batell:2009di,
	author         = "Batell, Brian and Pospelov, Maxim and Ritz, Adam",
	title          = "{Exploring Portals to a Hidden Sector Through Fixed
	Targets}",
	journal        = "Phys. Rev.",
	volume         = "D80",
	year           = "2009",
	pages          = "095024",
	doi            = "10.1103/PhysRevD.80.095024",
	eprint         = "0906.5614",
	archivePrefix  = "arXiv",
	primaryClass   = "hep-ph",
	SLACcitation   = "%%CITATION = ARXIV:0906.5614;%%"
}

@article{Essig:2010gu,
	author         = "Essig, Rouven and Harnik, Roni and Kaplan, Jared and
	Toro, Natalia",
	title          = "{Discovering New Light States at Neutrino Experiments}",
	journal        = "Phys. Rev.",
	volume         = "D82",
	year           = "2010",
	pages          = "113008",
	doi            = "10.1103/PhysRevD.82.113008",
	eprint         = "1008.0636",
	archivePrefix  = "arXiv",
	primaryClass   = "hep-ph",
	reportNumber   = "SLAC-PUB-14197, FERMILAB-PUB-10-274-T",
	SLACcitation   = "%%CITATION = ARXIV:1008.0636;%%"
}

@article{Lees:2014xha,
	author         = "Lees, J. P. and others",
	title          = "{Search for a Dark Photon in $e^+e^-$ Collisions at
	BaBar}",
	collaboration  = "BaBar",
	journal        = "Phys. Rev. Lett.",
	volume         = "113",
	year           = "2014",
	number         = "20",
	pages          = "201801",
	doi            = "10.1103/PhysRevLett.113.201801",
	eprint         = "1406.2980",
	archivePrefix  = "arXiv",
	primaryClass   = "hep-ex",
	reportNumber   = "BABAR-PUB-14-002, SLAC-PUB-15979",
	SLACcitation   = "%%CITATION = ARXIV:1406.2980;%%"
}

@article{TheBABAR:2016rlg,
	author         = "Lees, J. P. and others",
	title          = "{Search for a muonic dark force at BABAR}",
	collaboration  = "BaBar",
	journal        = "Phys. Rev.",
	volume         = "D94",
	year           = "2016",
	number         = "1",
	pages          = "011102",
	doi            = "10.1103/PhysRevD.94.011102",
	eprint         = "1606.03501",
	archivePrefix  = "arXiv",
	primaryClass   = "hep-ex",
	reportNumber   = "BABAR-PUB-16-003, SLAC-PUB-16549",
	SLACcitation   = "%%CITATION = ARXIV:1606.03501;%%"
}

@article{Harnik:2012ni,
	author         = "Harnik, Roni and Kopp, Joachim and Machado, Pedro A. N.",
	title          = "{Exploring nu Signals in Dark Matter Detectors}",
	journal        = "JCAP",
	volume         = "1207",
	year           = "2012",
	pages          = "026",
	doi            = "10.1088/1475-7516/2012/07/026",
	eprint         = "1202.6073",
	archivePrefix  = "arXiv",
	primaryClass   = "hep-ph",
	reportNumber   = "FERMILAB-PUB-12-048-T",
	SLACcitation   = "%%CITATION = ARXIV:1202.6073;%%"
}

@article{Adelberger:2006dh,
	author         = "Adelberger, E. G. and Heckel, Blayne R. and Hoedl, Seth
	A. and Hoyle, C. D. and Kapner, D. J. and Upadhye, A.",
	title          = "{Particle Physics Implications of a Recent Test of the
	Gravitational Inverse Sqaure Law}",
	journal        = "Phys. Rev. Lett.",
	volume         = "98",
	year           = "2007",
	pages          = "131104",
	doi            = "10.1103/PhysRevLett.98.131104",
	eprint         = "hep-ph/0611223",
	archivePrefix  = "arXiv",
	primaryClass   = "hep-ph",
	SLACcitation   = "%%CITATION = HEP-PH/0611223;%%"
}

@article{Schlamminger:2007ht,
	author         = "Schlamminger, Stephan and Choi, K. -Y. and Wagner, T. A.
	and Gundlach, J. H. and Adelberger, E. G.",
	title          = "{Test of the equivalence principle using a rotating
	torsion balance}",
	journal        = "Phys. Rev. Lett.",
	volume         = "100",
	year           = "2008",
	pages          = "041101",
	doi            = "10.1103/PhysRevLett.100.041101",
	eprint         = "0712.0607",
	archivePrefix  = "arXiv",
	primaryClass   = "gr-qc",
	SLACcitation   = "%%CITATION = ARXIV:0712.0607;%%"
}

@article{Choi:2019ixb,
	author = "Choi, Ki-Young and Kim, Jongkuk and Rott, Carsten",
	title = "{Constraining dark matter-neutrino interactions with IceCube-170922A}",
	eprint = "1903.03302",
	archivePrefix = "arXiv",
	primaryClass = "astro-ph.CO",
	doi = "10.1103/PhysRevD.99.083018",
	journal = "Phys. Rev. D",
	volume = "99",
	number = "8",
	pages = "083018",
	year = "2019"
}

@article{Khan:2017ygl,
	author = "Khan, Najimuddin",
	title = "{Neutrino mass and the Higgs portal dark matter in the ESSFSM}",
	eprint = "1707.07300",
	archivePrefix = "arXiv",
	primaryClass = "hep-ph",
	doi = "10.1155/2018/4809682",
	journal = "Adv. High Energy Phys.",
	volume = "2018",
	pages = "4809682",
	year = "2018"
}

@unpublished{Smirnov2020,
	title= {Neutrino and the Dark side of the Universe},
	author = {A. Y.  Smirnov},
	year = {2020},
	note= {3rd World Summit on Exploring
	the Dark Side of the Universe},
	URL= {https://indico.cern.ch/event/801461/contributions/3728174/},
}

@article{Adey:2018zwh,
	author = "Adey, D. and others",
	collaboration = "Daya Bay",
	title = "{Measurement of the Electron Antineutrino Oscillation with 1958 Days of Operation at Daya Bay}",
	eprint = "1809.02261",
	archivePrefix = "arXiv",
	primaryClass = "hep-ex",
	doi = "10.1103/PhysRevLett.121.241805",
	journal = "Phys. Rev. Lett.",
	volume = "121",
	number = "24",
	pages = "241805",
	year = "2018"
}

@article{Bak:2018ydk,
	author = "Bak, G. and others",
	collaboration = "RENO",
	title = "{Measurement of Reactor Antineutrino Oscillation Amplitude and Frequency at RENO}",
	eprint = "1806.00248",
	archivePrefix = "arXiv",
	primaryClass = "hep-ex",
	doi = "10.1103/PhysRevLett.121.201801",
	journal = "Phys. Rev. Lett.",
	volume = "121",
	number = "20",
	pages = "201801",
	year = "2018"
}

@article{DoubleChooz:2019qbj,
	author = "de Kerret, H. and others",
	collaboration = "Double Chooz",
	title = "{Double Chooz \ensuremath{\theta}$_{13}$ measurement via total neutron capture detection}",
	eprint = "1901.09445",
	archivePrefix = "arXiv",
	primaryClass = "hep-ex",
	doi = "10.1038/s41567-020-0831-y",
	journal = "Nature Phys.",
	volume = "16",
	number = "5",
	pages = "558--564",
	year = "2020"
}

@article{Abe:2020vot,
	author = "Abe, K. and others",
	collaboration = "T2K",
	title = "{Measurement of the charged-current electron (anti-)neutrino inclusive cross-sections at the T2K off-axis near detector ND280}",
	eprint = "2002.11986",
	archivePrefix = "arXiv",
	primaryClass = "hep-ex",
	doi = "10.1007/JHEP10(2020)114",
	journal = "JHEP",
	volume = "10",
	pages = "114",
	year = "2020"
}

@article{Abe:2019vii,
	author = "Abe, K. and others",
	collaboration = "T2K",
	title = "{Constraint on the matter\textendash{}antimatter symmetry-violating phase in neutrino oscillations}",
	eprint = "1910.03887",
	archivePrefix = "arXiv",
	primaryClass = "hep-ex",
	doi = "10.1038/s41586-020-2177-0",
	journal = "Nature",
	volume = "580",
	number = "7803",
	pages = "339--344",
	year = "2020",
	note = "[Erratum: Nature 583, E16 (2020)]"
}

@article{Acero:2019ksn,
	author = "Acero, M. A. and others",
	collaboration = "NOvA",
	title = "{First Measurement of Neutrino Oscillation Parameters using Neutrinos and Antineutrinos by NOvA}",
	eprint = "1906.04907",
	archivePrefix = "arXiv",
	primaryClass = "hep-ex",
	reportNumber = "FERMILAB-PUB-19-272-ND",
	doi = "10.1103/PhysRevLett.123.151803",
	journal = "Phys. Rev. Lett.",
	volume = "123",
	number = "15",
	pages = "151803",
	year = "2019"
}

@article{Adamson:2020ypy,
	author = "Adamson, P. and others",
	collaboration = "MINOS+",
	title = "{Precision Constraints for Three-Flavor Neutrino Oscillations from the Full MINOS+ and MINOS Dataset}",
	eprint = "2006.15208",
	archivePrefix = "arXiv",
	primaryClass = "hep-ex",
	reportNumber = "FERMILAB-PUB-20-253-ND",
	doi = "10.1103/PhysRevLett.125.131802",
	journal = "Phys. Rev. Lett.",
	volume = "125",
	number = "13",
	pages = "131802",
	year = "2020"
}

@article{Aartsen:2019tjl,
	author = "Aartsen, M. G. and others",
	collaboration = "IceCube",
	title = "{Measurement of Atmospheric Tau Neutrino Appearance with IceCube DeepCore}",
	eprint = "1901.05366",
	archivePrefix = "arXiv",
	primaryClass = "hep-ex",
	doi = "10.1103/PhysRevD.99.032007",
	journal = "Phys. Rev. D",
	volume = "99",
	number = "3",
	pages = "032007",
	year = "2019"
}

@article{Albert:2018mnz,
	author = "Albert, A. and others",
	collaboration = "ANTARES",
	title = "{Measuring the atmospheric neutrino oscillation parameters and constraining the 3+1 neutrino model with ten years of ANTARES data}",
	eprint = "1812.08650",
	archivePrefix = "arXiv",
	primaryClass = "hep-ex",
	doi = "10.1007/JHEP06(2019)113",
	journal = "JHEP",
	volume = "06",
	pages = "113",
	year = "2019"
}

\end{document}